\documentclass[aps,pre,showpacs,twocolumn]{revtex4-1} 
\usepackage{amsmath}
\usepackage{graphicx}

\newcommand{\dbar} {\ensuremath{\,\mathchar'26\mkern-12mu d}}

\newcommand{\Micro}{\mathbf{x}}

\newcommand{\Bath}{\mathbf{y}}
\newcommand{\Time}{t}

\newcommand{\EndTime}{\tau}

\newcommand{\Rev}[1]{\hat{#1}}
\newcommand{\RevState}[1]{#1^*}
\newcommand{\TrajProb}{p(\Micro_0^\EndTime |\Micro_0)}

\newcommand{\TrajProbRevEq}{p(\Rev{\Micro}_0^\EndTime|\RevState{\Micro}_\EndTime)}
\newcommand{\TrajProbControl}{p(\Micro_0^\EndTime|\Micro_0, \Control_0^\EndTime)}
\newcommand{\TrajProbRevControl}{p(\Rev{\Micro}_0^\EndTime|\RevState{\Micro}_\EndTime, \Rev{\Control}_0^\EndTime)}

\newcommand{\EntropyExt}{S_e}
\newcommand{\EntropyInt}{s}

\newcommand{\Heat}{Q}

\newcommand{\Number}{n}
\newcommand{\Temp}{T}
\newcommand{\Chem}{\mu}
\newcommand{\Boltz}{k_B}

\newcommand{\Force}{f}

\newcommand{\Work}{W}
\newcommand{\WorkGen}{\mathcal{W}}
\newcommand{\Free}{F}
\newcommand{\FreeGen}{\mathcal{F}}

\newcommand{\Control}{\lambda}
\newcommand{\Func}[1]{\mathcal{F}[#1]}
\newcommand{\Measure}[1]{\mathcal{D}[#1]}

\newcommand{\Energy}{U}
\newcommand{\Htot}{H_{\rm tot}}
\newcommand{\Hsys}{H_{\rm sys}}
\newcommand{\Hint}{h_{\rm int}}
\newcommand{\Hbath}{H_{\rm env}}
\newcommand{\NumberOfStates}{\Omega}
\newcommand{\VolTraj}{\NumberOfStates(\Micro_0^\EndTime)}
\newcommand{\VolTrajRev}{\RevState{\NumberOfStates}(\Rev{\Micro}_0^\EndTime)}
\newcommand{\VolStart}{\NumberOfStates(\Energy-\Hsys(\Micro_0,\Control_0))}
\newcommand{\VolEnd}{\NumberOfStates(\Energy+\Work-\Hsys(\Micro_\EndTime,\Control_\EndTime))}

\newcommand{\MicroProbFwd}{p_0(\Micro_0)}
\newcommand{\MicroProbRev}{p_\EndTime(\Micro_\EndTime)}

\newcommand{\MacroI}{{\rm I}}
\newcommand{\MacroII}{{\rm II}}
\newcommand{\MacroIII}{{\rm III}}

\begin{document}

\title{Limits of Predictions in Thermodynamic Systems: A Review}
\author{Robert Marsland III}
\affiliation{Department of Physics, Boston University, 590 Comm. Ave., Boston, Massachusetts 02215, USA}
\email{marsland@bu.edu}
\author{Jeremy England}
\affiliation{Physics of Living Systems, Massachusetts Institute of Technology, 400 Technology Square, Cambridge, Massachusetts 02139, USA}
\date{\today}


\begin{abstract}
	The past twenty years have seen a resurgence of interest in nonequilibrium thermodynamics, thanks to advances in the theory of stochastic processes and in their thermodynamic interpretation. Fluctuation theorems provide fundamental constraints on the dynamics of systems arbitrarily far from thermal equilibrium. Thermodynamic uncertainty relations bound the dissipative cost of precision in a wide variety of processes. Concepts of excess work and excess heat provide the basis for a complete thermodynamics of nonequilibrium steady states, including generalized Clausius relations and thermodynamic potentials. But these general results carry their own limitations: fluctuation theorems involve exponential averages that can depend sensitively on unobservably rare trajectories; steady-state thermodynamics makes use of a dual dynamics that lacks any direct physical interpretation. This review aims to present these central results of contemporary nonequilibrium thermodynamics in such a way that the power of each claim for making physical predictions can be clearly assessed, using examples from current topics in soft matter and biophysics. 
\end{abstract}

\maketitle

\section{Introduction}
Modern society is built on the transformation of heat into mechanical work. The task of perfecting the heat engine was made possible by theoretical insights into fundamental bounds on its efficiency. By calculating the maximum amount of work extractable from a given quantity of heat by an engine operating between two thermal reservoirs of fixed temperature, Sadi Carnot was able to show that state-of-the-art steam engines in 1824 could still be significantly improved \cite{Carnot}. By determining the kinds of processes that saturate the bound, he was able to offer practical suggestions for making these improvements. 

Biotechnology and biologically inspired design have opened up new engineering challenges that push the boundaries of thermodynamics. Living systems use heat to accomplish a wide variety of tasks in addition to performing mechanical work, such as replicating complex structures with high fidelity \cite{Murugan2012,Murugan2014,Rao2015}, maintaining robust clocks \cite{Cao2015,Barato2016}, and sensing the state of their environment \cite{Lan2012,Lang2014,Sartori2014}. Most of these processes are intrinsically nonequilibrium, in the sense that they cannot be represented even approximately as a sequence of states of thermal equilibrium. Since they never approach the limit of zero entropy change, such processes are left unconstrained by classical thermodynamic theory. Similar issues arise in contemporary soft matter physics, particularly with the advent of active matter systems that constantly transduce chemical energy into mechanical force on microscopic scales \cite{Sanchez2012,Chaikin2013,Marchetti2013}.

In the 1950's-1970's, the theory of Linear Irreversible Thermodynamics was developed to analyze chemical reactions and various transport problems near thermal equilibrium from a thermodynamic perspective (cf. \cite{deGroot} for a thorough introduction). Many hoped that these results could be developed into a general theory of nonequilibrium thermodynamics, capable of making universal claims analogous to the Carnot bound on engine efficiency \cite{Prigogine1955,Oono1998}. Ilya Prigogine's Principle of Minimum Entropy Production initially seemed like a promising starting point for constructing a more comprehensive theory, but counterexamples were soon identified that revealed its restricted range of validity (cf. \cite[p. 100ff.]{Prigogine1955},\cite{Landauer1975}).

The past twenty years have seen a resurgence of interest in nonequilibrium thermodynamics, based on advances in the theory of stochastic processes and in their thermodynamic analysis. Stochastic Thermodynamics has matured into a systematic theory of nonequilibrium processes, in which the analogies to thermal equilibrium that Prigogine and others were searching for can be mathematically defined \cite{Seifert2012,VandenBroeck2010}. But these general results carry their own limits: far from equilibrium, they demand precise knowledge of the probability distributions of the relevant quantities, which are often experimentally inaccessible \cite{Jarzynski2006}, or they invoke a ``dual'' dynamics that lacks any direct physical interpretation \cite{Sasa2014}. This review aims to present a selection of central results from contemporary nonequilibrium thermodynamics in such a way that the power of each claim for making physical predictions can be clearly assessed. 

After setting up our notation and theoretical framework in Section \ref{sec:micro}, we proceed in Section \ref{sec:shannon} to consider the immediate consequences of imposing a consistent thermodynamic interpretation on a stochastic model. These include fluctuation theorems, the role of Shannon entropy, and the relationship of entropy to transition rate ratios at a coarse-grained level. We illustrate these ideas and their potential applications by reviewing a recent effort to obtain general design principles for nonequilibrium self-assembly from thermodynamic considerations.

The practical utility of the results of Section \ref{sec:shannon} is limited by their dependence on information about exponentially rare fluctuations, which becomes important far from equilibrium. Recent advances in the study of such fluctuations in the field of Large Deviation Theory have significantly pushed back this limit. In Section \ref{sec:uncertainty}, we discuss the assumptions and consequences of a particularly powerful result that relates these rare fluctuations to the small fluctuations that could actually be observed in an experiment \cite{Gingrich2016,Pietzonka2016c}. This theorem leads to a ``thermodynamic uncertainty relation,'' which places a limit on the allowed dynamical precision of a nonequilibrium process based on its rate of entropy production \cite{Barato2015}.

If we now imagine perturbing a driven system -- compressing an active material, for instance -- many analogies to classical thermodynamics suggest themselves. In equilibrium, we could have computed the force resisting the perturbation by taking a derivative of free energy; we could have predicted typical values of observables by minimizing the free energy; and we could have constrained the minimum work required to change they state of the system with the difference between initial and final free energies. In Section \ref{sec:excess}, we discuss the extent to which these three kinds of thermodynamic predictions can be generalized to transitions between nonequilibrium steady states.

We believe that this selection of topics will provide a helpful unifying perspective on some of the most important themes in contemporary nonequilibrium thermodynamics, particularly for readers interested in biophysical or soft-matter applications. But a number of current lines of inquiry have been left outside the scope of this review. Stochastic Thermodynamics has shed new light on the original thermodynamic problem of heat engine performance (cf. \cite{Brandner2015,Raz2016}), consolidating the considerable progress made in Finite-Time Thermodynamics throughout the second half of the 20th century \cite{Bjarne2011,Ouerdane2015}. This new perspective has also helped elucidate the nature of fluctuation-response relations, showing how the traditional formulations can be generalized to far-from-equilibrium scenarios \cite{Marconi2008,Baiesi2013}. A comprehensive review of Stochastic Thermodynamics from 2012 summarizes the development of both of these themes up through that year \cite{Seifert2012}. 

\section{Microscopic reversibility connects entropy to dynamics}
\label{sec:micro}
In thermal equilibrium, the principle of detailed balance relates transition rates between system states $\Micro$ to the energies of those states $\Energy(\Micro)$, by requiring that all probability flux vanish in the Boltzmann distribution. We can express this requirement in terms of the probability that the system passes through a continuous sequence of states $\Micro_\Time$ during a window of duration $\EndTime$, starting from state $\Micro_0$ at time $t=0$. We denote the whole trajectory by $\Micro_0^\EndTime$, and the time-reversed version of this trajectory by $\Rev{\Micro}_0^\EndTime$. At time $\Time$ after initialization, a system following the time-reversed trajectory will be found in state $\RevState{\Micro}_{\EndTime-\Time}$, where $\RevState{\Micro}$ is the state obtained from $\Micro$ by reversing the signs of all momentum degrees of freedom that may be contained in that vector. 

Detailed balance requires that the ratio of forward and reverse trajectory probabilities $\TrajProb$ and $\TrajProbRevEq$ conditioned on their respective initial states must be equal to the ratio of Boltzmann weights:
\begin{align}
\label{eq:detailed}
\frac{\TrajProbRevEq}{\TrajProb} = e^{\frac{1}{\Boltz\Temp}[\Energy(\Micro_\EndTime) - \Energy(\Micro_0)]}.
\end{align}

The cornerstone of contemporary nonequilibrium statistical mechanics is a generalization of Equation (\ref{eq:detailed}) known as ``local detailed balance'' or ``microscopic reversibility,'' which applies to driven systems arbitrarily far from thermal equilibrium. Since this principle is so fundamental, it is important to understand its physical basis and its range of validity before proceeding to more specific results.

\subsection{Stochastic processes from classical mechanics}
 Consider an isolated chunk of classical matter with Hamiltonian $\Htot$, whose state is described by a set of $\mathcal{N}$ coordinates $q_i$ with conjugate momenta $p_i$. The dynamics are given by Hamilton's equations  
\begin{align}
\label{eq:Hamil}
\dot{q_i} &= \frac{\partial\Htot}{\partial p_i}\\
\dot{p_i} &= -\frac{\partial\Htot}{\partial q_i}\nonumber.
\end{align}
We will keep track of some subset of the $\mathcal{N}$ degrees of freedom, which we call the ``system'' and denote by $\Micro$, and refer to the rest of the degrees of freedom $\Bath$ as the ``environment.'' We can now split the Hamiltonian into three parts: one that depends on the system degrees of freedom alone, one that depends on the environment alone, and one that couples the two sets together:
\begin{align}
\Htot(\Micro,\Bath,\Time) = \Hsys(\Micro,\Control_\Time) + \Hbath(\Bath) + \Hint(\Micro,\Bath)
\end{align}
We have allowed for an explicit time-dependence in the Hamiltonian via a control parameter $\Control$ (e.g., the position of the piston in the classic example of gas compression), which can be used to do work on the system according to a pre-defined protocol $\Control_0^\EndTime$ and drive it out of equilibrium. 

\begin{figure}
	\centering
	\includegraphics[width=8cm]{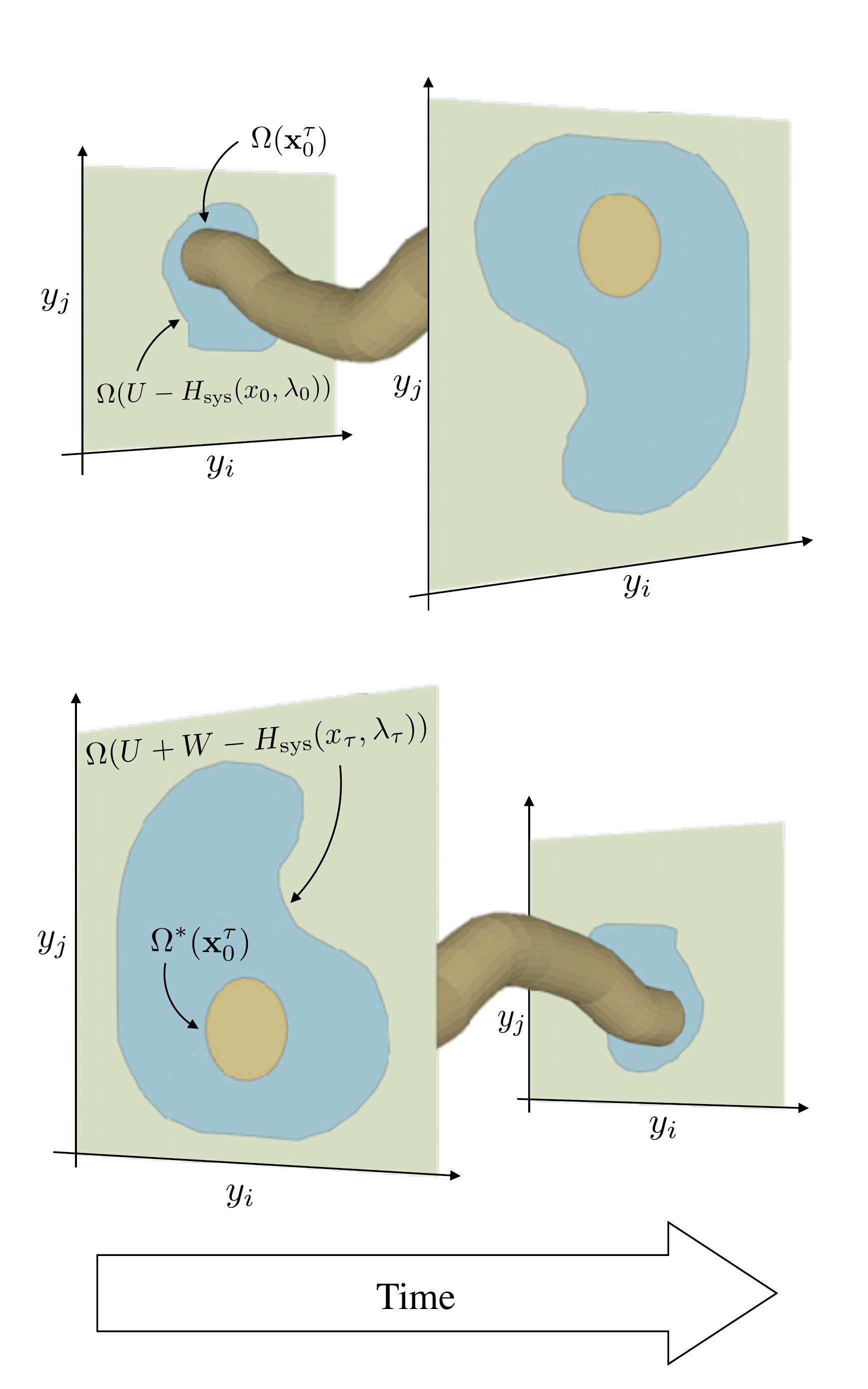}
	\caption{Color online. Time-reversal symmetry requires that the environmental phase space volumes giving rise to forward and reverse trajectories are identical. The planes represent the phase space of the environment, and the tubes represent the sets of environmental trajectories compatible with a given system trajectory $\Micro_0^\EndTime$ and its time-reverse $\Rev{\Micro}_0^\EndTime$, respectively. The conservation of the Liouville measure implies that the beginning and end of the tube occupy the same phase space volume, and time-reversal symmetry requires that the initial conditions for the reverse tube are simply the momentum-reversed versions of the final conditions for the forward tube.}
	\label{fig:phasespace}
\end{figure}

We now fix the initial condition of the system $\Micro_0$, and choose a random initial condition $\Bath_0$ for the environment from a uniform distribution over states with total energy $\Htot = \Energy$. Since the dynamics of the whole setup are deterministic, the trajectory $\Micro_0^\EndTime$ is fully determined by this initial random choice. We can collect all the $\Bath_0$'s that produce trajectories within a small window around a given $\Micro_0^\EndTime$, and denote the phase space volume occupied by this set of states as $\VolTraj$. The probability of observing a trajectory in the window is then given by the ratio of this volume to the total phase space volume $\VolStart$ available to the environment at time $\Time = 0$:
\begin{align}
\label{eq:probf}
\TrajProbControl = \frac{\VolTraj}{\VolStart}.
\end{align}

The energy of the environment changes over the course of the trajectory, as heat flows in and out of the system. At the end of the trajectory, it is equal to $E + \Work - \Hsys(\Micro_\EndTime, \Control_\EndTime)$, where the work $\Work$ done by manipulation of $\Control$ is 
\begin{align}
\label{eq:Work0}
\Work = \int_0^\EndTime \frac{\partial \Hsys(\Micro_\Time,\Control_\Time)}{\partial \Control}\dot{\Control_\Time} d\Time.
\end{align}
Now we can compute the probability of seeing the reverse trajectory $\Rev{\Micro}_0^\EndTime$ if we reverse the momenta, choose the environment conditions from this new energy surface, and run the protocol $\Control_\Time$ in reverse:
\begin{align}
\label{eq:probr}
\TrajProbRevControl = \frac{\VolTrajRev}{\VolEnd}.
\end{align}

As illustrated in Figure \ref{fig:phasespace}, the time-reversibility of the Hamiltonian dynamics (\ref{eq:Hamil}) can be expressed as the statement
\begin{align}
\label{eq:volsym}
\VolTraj = \VolTrajRev.
\end{align}
This relation expresses the fact that every trajectory $(\Micro_0^\EndTime,\Bath_0^\EndTime)$ satisfying the equations of motion (\ref{eq:Hamil}) has a reverse trajectory $(\Rev{\Micro}_0^\EndTime,\Rev{\Bath}_0^\EndTime)$ that also satisfies these equations. (It also assumes that phase space volumes $\NumberOfStates = \int \prod_i dp_i \,dq_i$ are measured with the Liouville measure $\prod_i dp_i\,dq_i$, which is conserved by the equations of motion.) Note that Equation (\ref{eq:volsym}) remains true even in the presence of magnetic fields, as long as the signs of these fields are reversed for the purposes of calculating $\VolTrajRev$. 

Combining Equations (\ref{eq:probf}), (\ref{eq:probr}) and (\ref{eq:volsym}) leads immediately to
\begin{align}
\frac{\TrajProbRevControl}{\TrajProbControl} = \frac{\VolStart}{\VolEnd}.
\end{align}
We can now use Boltzmann's microcanonical definition of entropy to express the right hand side in terms of the thermodynamic entropy $\EntropyExt = \Boltz \ln \NumberOfStates$ of the environment:
\begin{align}
\label{eq:MicroRev}
\frac{\TrajProbRevControl}{\TrajProbControl} =  e^{-\Delta \EntropyExt/\Boltz}.
\end{align}
Equation (\ref{eq:MicroRev}) is known as the ``microscopic reversibility relation,'' since it encodes the time-reversibility of the full microscopic dynamics in a coarse-grained stochastic dynamics \cite{Crooks1999,Esposito2010}.

\subsection{Entropy and Heat}
The microscopic reversibility relation (\ref{eq:MicroRev}) has a remarkably wide range of validity, although the above derivation has been made somewhat more restrictive for the sake of clarity and simplicity. In particular, we have assumed that the environment is initialized in thermal equilibrium, so that the initial environmental states can be sampled from a uniform distribution over a constant-energy surface. The microcanonical definition of temperature $1/\Temp = \partial S/\partial \Energy$ then yields
\begin{align}
\label{eq:heat}
\Delta\EntropyExt = \frac{\Heat}{\Temp}
\end{align}
where $\Heat = \Work - \Delta \Hsys$ is the amount of energy added to the environment over the course of the given trajectory.

\begin{figure}
	\centering
	\includegraphics[width=8.5cm]{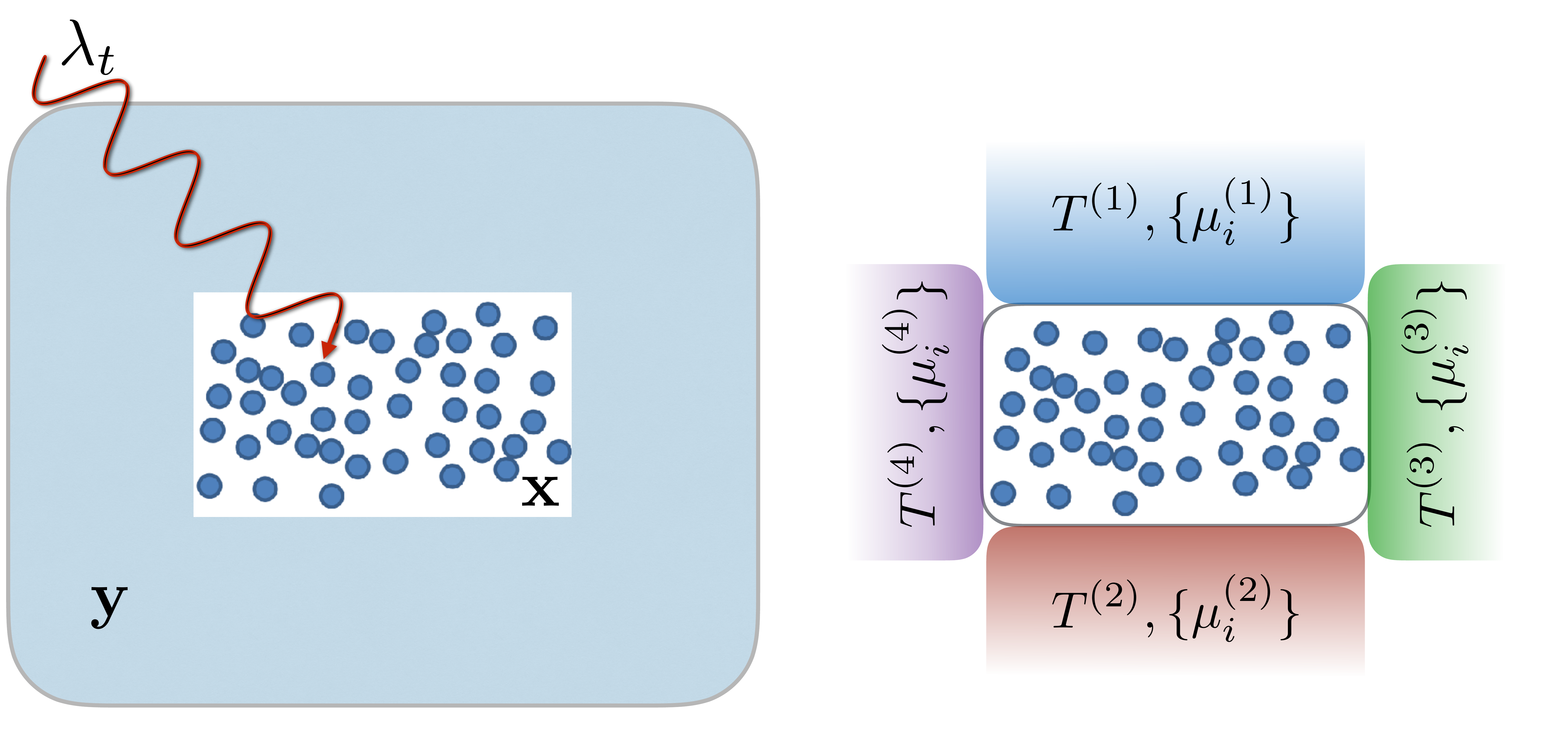}
	\caption{Color online. Left: The $\mathcal{N}$ degrees of freedom in a piece of isolated matter are partitioned into two sets, a system with microstate $\Micro$ and an environment with microstate $\Bath$. The system is illustrated here as discrete particles, since we keep track of the full trajectories $\Micro_0^\EndTime$. The environment is a solid color, because we integrate it out to obtain an effective stochastic dynamics for $\Micro$. Work can be done on the system by externally imposed variations in a set of parameters $\Control_t$, which affect only the system part of the Hamiltonian $\Hsys$. Right: We consider situations where the environment can be modeled as a set of ideal thermal and chemical reservoirs with temperatures $\Temp^{(\alpha)}$ and chemical potentials $\Chem_i^{(\alpha)}$.}
	\label{fig:SysEnv}
\end{figure}

But the derivation can be straightforwardly generalized to the case where different sectors of the environment are initially equilibrated separately with different temperatures $\Temp^{(\alpha)}$ and are brought into contact with the system at time $t = 0$. The result also remains valid when particle exchange between system and environment is allowed, although this is more complicated to set up notationally (see \cite{Komatsu2008} for derivation of a related expression allowing for particle exchange). If we include these possibilities, illustrated in Figure \ref{fig:SysEnv}, we obtain a more general expression for the entropy change (cf. \cite{Polettini2016b}):
\begin{align}
\label{eq:entropyext}
\Delta \EntropyExt[\Micro_0^\EndTime ]= \sum_\alpha \frac{1}{\Temp^{(\alpha)}} \left( \Heat^{(\alpha)}[\Micro_0^\EndTime] - \sum_i\Chem^{(\alpha)}_i\Delta\Number_i^{(\alpha)}[\Micro_0^\EndTime]\right).
\end{align}
where $\Heat^{(\alpha)}$ and $\Delta \Number^{(\alpha)}_i$ are the heat and number of particles of type $i$, respectively, delivered to sectors of the environment with temperatures $\Temp^{(\alpha)}$ and chemical potentials $\mu^{(\alpha)}_i$. 

In subsequent sections, it will be convenient to split $\Delta\EntropyExt$ into an equilibrium contribution $-\Delta\Hsys/\Temp$ and a nonequilibrium correction $\WorkGen$ that can be thought of as a generalized work:
\begin{align}
\label{eq:FirstLawGen}
\Temp \Delta \EntropyExt = \WorkGen - \Delta\Hsys.
\end{align}
The generalized work can be written explicitly as
\begin{align}
\label{eq:workgen}
\WorkGen &=\Work +\Temp \Delta\EntropyExt - \sum_\alpha \Heat^{(\alpha)},
\end{align}
and reduces to the ordinary mechanical work $\Work$ performed through the control parameters $\Control_t$ in the case of an isothermal process with no particle exchange. This splitting highlights the fact that systems can be kept out of equilibrium by thermal or chemical ``work'' even if all the control parameters $\Control$ are held constant. In such cases, the system relaxes to a nonequilibrium steady state (in the limit of infinitely large reservoirs), which will be the main subject of Section \ref{sec:excess}. 

Since the Schr\"{o}dinger Equation is time-reversible, a version of Equation (\ref{eq:MicroRev}) can also been obtained for quantum systems. In this case the system trajectory cannot be observed without interfering with the dynamics, however, and one has to consider the environment trajectory instead \cite{Crooks2008,Horowitz2013}. The consequences of microscopic reversibility that we will discuss in Section \ref{sec:shannon} do not involve observation of fully-detailed system trajectories, and are thus easier to generalize to the quantum case \cite{Campisi2011}.

In summary, the microscopic reversibility relation (\ref{eq:MicroRev}) depends on two basic assumptions: (a) the intrinsic dynamics of the system are time reversible, and (b) the environment is initialized in thermal equilibrium, or is partitioned into non-interacting sectors that are each separately initialized in thermal equilibrium.


\subsection{Coarse-grained models}
\label{sec:models}
A nonequilibrium steady state can also be set up by other means, including stationary force fields with non-zero curl and externally imposed flow fields. Physically, all these forms of driving are ultimately reducible to the above framework. Experimental realizations of non-conservative force fields, for example, are actually created by rapidly varying some control parameters \cite{Blickle2007}. But it is convenient to model such a system by abstracting from the source of the flow field or effective force field, and these models can be numerically simulated without adding more information about the underlying oscillations or gradients. 

Furthermore, many systems of interest in soft matter and biology consist of very small particles in aqueous solution. It is often a good approximation to let the momentum degrees of freedom and the microstate of the water molecules instantaneously relax to thermal equilibrium, taking the spatial location of the particles of interest to fully specify the system state $\Micro$. The dynamics of $\Micro$ are then described by an overdamped Langevin equation, which makes the particle's velocity proportional to the applied force and includes a random force due to the hidden degrees of freedom (cf. \cite{Seifert2012,Gardiner}). 

Chemically driven systems are usually represented with a similar sort of coarse-grained dynamics, with ``fast'' degrees of freedom rapidly reaching the Boltzmann distribution within a subset of microstates specified by ``slow'' variables. The slow variables in a chemical system refer to discrete metastable states, which could contain a collection of many similar protein conformations in the same free energy well, or specify the concentrations of several molecular species in a well-mixed volume. 

These coarse-grained models are no longer guaranteed to satisfy Equation (\ref{eq:MicroRev}), and it is necessary to directly verify that the equation does indeed hold under the natural definitions of heat and work in the model. This is straightforward for diffusion processes driven by non-conservative forces (cf. \cite{Sasa2014}) or by external flows \cite{Speck2008a}, which satisfy (\ref{eq:MicroRev}) when the heat is obtained from the natural definition of work as force times distance and of energy as the integral of the conservative part of the force. 

Chemical systems require more care. If the dynamics on the set of discrete states $\Micro$ is modeled as a Markovian stochastic process, and the assumption of separation of timescales holds, one obtains a modified form of (\ref{eq:MicroRev}). The heat $Q$ in Equation (\ref{eq:entropyext}) is replaced by the free energy difference $-\Delta \Free$ between the final and initial coarse-grained states. In general, this free energy is given by $\Free(\Micro) = -\Boltz \Temp \ln Z_\Micro$ where $Z_\Micro$ is the sum of the Boltzmann weight $e^{-\Energy/\Boltz\Temp}$ over all the hidden degrees of freedom within the coarse-grained state $\Micro$ \cite{Seifert2012}. When $\Micro$ is a vector of chemical concentrations in a well-mixed volume, the free energy contains a part due to the intrinsic properties of the molecule, and part due to the sum over the possible spatial configurations of particles within the volume. In this case, the contribution to the free energy from each chemical species is equivalent to a chemical potential $\mu_i = \mu_{0,i} + \Boltz\Temp \ln c_i$ \cite{Schnakenberg1976}.

In the limit of large system size, where the concentrations are naturally treated as continuous variables, chemical reaction dynamics are often modeled with a stochastic differential equation known as the Chemical Langevin Equation (CLE). Near equilibrium, the CLE still obeys Equation (\ref{eq:MicroRev}). But once the free energy change per reaction becomes of order $\Boltz\Temp$, the ratio of reverse to forward trajectory probabilities no longer corresponds directly to any thermodynamic quantity \cite{Horowitz2015}. 

\section{Consequences of microscopic reversibility}
\label{sec:shannon}
The basic insights of equilibrium thermodynamics are founded on the Clausius relation
\begin{align}
\label{eq:Clausius}
\Delta S \geq - \frac{\Heat}{\Temp}
\end{align}
which relates the change $\Delta S$ in the entropy of a system to the heat $\Heat$ exhausted into an ideal thermal reservoir at temperature $\Temp$. More generally, one can replace $\Heat/\Temp$ by $\Delta\EntropyExt$, which could include contributions from particle exchange and multiple thermal reservoirs. 

The entropy of equilibrium thermodynamics is only defined for equilibrium states. To obtain thermodynamic constraints on transitions between nonequilibrium states, it is necessary to define a generalized entropy that still obeys Equation (\ref{eq:Clausius}). The Gibbs/Shannon entropy 
\begin{align}
S = -\Boltz \int d\Micro \,p(\Micro)\ln p(\Micro)
\end{align}
is naturally generalizable to nonequilibrium states, since it remains well-defined even when $p(\Micro)$ is not the Boltzmann distribution. J. Schnakenberg argued in the 1970's that Equation (\ref{eq:Clausius}) should hold for this choice of the entropy in a broad class of Markovian stochastic processes on a finite set of states \cite{Schnakenberg1976}. 

In this section, we review how a generalized version of Schnakenberg's result follows from microscopic reversibility (\ref{eq:MicroRev}), thus confirming that the Shannon entropy preserves the Clausius inequality away from thermal equilibrium. We then point out a further refinement that relates the average total entropy change to macroscopic irreversibility, placing a tighter constraint on processes that necessarily involve positive total entropy production. We end the section by discussing design principles for nonequilibrium self-assembly that can be derived from these results, and commenting on their range of applicability.

\subsection{Fluctuation Theorem for Total Entropy Change}
The full trajectories $\Micro_0^\EndTime$ that appear in Equation (\ref{eq:MicroRev}) are usually not directly observable. To obtain predictions from this symmetry, we need to integrate out the dependence on these microscopic details. As we will see, integrating Equation (\ref{eq:MicroRev}) over all trajectories leads directly to an integral fluctuation theorem (FT). This theorem involves all moments of the probability distribution for total entropy change. In this section, we will be concerned with the resulting inequality for the first moment, which is a form of the Clausius inequality. By retaining the information about fluctuations that is discarded in this derivation, one can obtain additional predictions, including the Fluctuation-Dissipation Theorem of classical linear response theory (cf. \cite[Ch. 5]{CrooksThesis}).

Multiplying $\TrajProbControl$ by any normalized probability distribution $\MicroProbFwd$ generates a normalized distribution $\mathcal{P}(\Micro_0^\EndTime|\Control_0^\EndTime)$ over the whole trajectory space. Likewise, multiplying $\TrajProbRevControl$ by some $\MicroProbRev$ generates another normalized distribution $\mathcal{P}(\Rev{\Micro}_0^\EndTime|\Rev{\Control}_0^\EndTime)$. It follows immediately from Equation (\ref{eq:MicroRev}) and the normalization of these distributions that 
\begin{align}
\label{eq:fluc1}
\langle e^{-\frac{\Delta\EntropyExt}{\Boltz} + \ln \frac{\MicroProbRev}{\MicroProbFwd}}\rangle = 1
\end{align}
where $\langle \Func{\Micro_0^\EndTime} \rangle = \int \Measure{\Micro_0^\EndTime} \Func{\Micro_0^\EndTime}\mathcal{P}(\Micro_0^\EndTime|\Control_0^\EndTime)$ is the average value of the trajectory functional $\Func{\Micro_0^\EndTime}$ over the forward trajectory distribution $\mathcal{P}(\Micro_0^\EndTime|\Control_0^\EndTime)$. 

If $\MicroProbRev$ is chosen as the time-evolved version of $\MicroProbFwd$, then Equation (\ref{eq:fluc1}) can be written in the more suggestive form
\begin{align}
\label{eq:fluc2}
\langle e^{-\frac{1}{\Boltz}(\Delta\EntropyExt + \Delta\EntropyInt)}\rangle = 1
\end{align}
using the stochastic entropy
\begin{align}
\label{eq:entropyint}
\Delta \EntropyInt = -\Boltz \ln \MicroProbRev + \Boltz\ln \MicroProbFwd.
\end{align}
The average of this quantity over the forward trajectory ensemble gives the change in Shannon entropy of the system over the course of its forward evolution:
\begin{align}
\label{eq:ShannonChange}
\langle \Delta \EntropyInt \rangle =  -\Boltz &\int d\Micro \,\MicroProbRev \ln \MicroProbRev \nonumber\\
&+ \Boltz \int d\Micro \,\MicroProbFwd\ln \MicroProbFwd.
\end{align}
Equation (\ref{eq:fluc2}) is known as the integral fluctuation theorem for the total entropy change, and was first obtained in 2005 by Udo Seifert (who used it to justify the definition (\ref{eq:entropyint}) of the stochastic entropy) \cite{Seifert2005}. This is a very powerful result, and almost everything we will discuss in the remainder of this review follows from it or from some closely related variant.

Using the fact that $e^x > 1+x$, we recover the Clausius inequality
\begin{align}
\label{eq:StochClausius}
\langle \Delta\EntropyInt \rangle \geq - \langle \Delta\EntropyExt \rangle.
\end{align}
This confirms Schnakenberg's thesis for any system obeying microscopic reversibility (\ref{eq:MicroRev}), including situations beyond the scope of his original paper where the dynamical rules change in time due to variations in a control parameter $\Control_t$.

Equations (\ref{eq:fluc2}) and (\ref{eq:StochClausius}) can be rewritten in terms of the thermodynamic work $\WorkGen$ defined in Equation (\ref{eq:FirstLawGen}), which will be particularly useful in Section \ref{sec:excess} when discussing transitions between nonequilibrium steady states. The fluctuation theorem becomes
\begin{align}
\label{eq:JarGen}
\langle e^{-\frac{1}{\Boltz\Temp}(\WorkGen-\Delta \Energy + \Temp \Delta \EntropyInt)}\rangle = 1,
\end{align}
This reduces to the Jarzynski Equality when the initial and final distributions are chosen as Boltzmann distributions, which turns $\Energy - \Temp\EntropyInt$ into the equilibrium free energy $F$ \cite{Jarzynski1997}. In general, this fluctuation theorem implies the isothermal form of the Clausius inequality:
\begin{align}
\label{eq:StochClausiusWork}
 \langle \WorkGen\rangle \geq  \Delta \FreeGen
\end{align}
where the nonequilibrium free energy is
\begin{align}
\label{eq:freegen}
\FreeGen =  \Energy - \Temp \langle\EntropyInt\rangle.
\end{align}

\subsection{Statistical Irreversibility in Macroscopic Transitions}
Our starting point in Equation (\ref{eq:MicroRev}) quantified the statistical irreversibility of a microscopic trajectory and related that quantity to the entropy change. By modifying a few steps in the above derivation of the integral fluctuation theorem, one can preserve more of this information during the integration over trajectories, bounding the entropy change with a positive number that quantifies the irreversibility of a macroscopic transition \cite{Kawai2007,Gomez-Marin2008a,Horowitz2009,England2013}. 

One way of doing this is to categorize the microstates according to some empirical criterion \cite{England2013}. This partitions the phase space into regions that we label $\MacroI,\MacroII,\MacroIII,\dots$. We can study the statistics of transitions between a pair of regions $\MacroI$ and $\MacroII$ by setting $\MicroProbFwd = 0$ everywhere outside of region $\MacroI$ and $\MicroProbRev = 0$ everywhere outside of region $\MacroII$. For simplicity, we will require the distributions within each region to satisfy $p(\RevState{\Micro}) = p(\Micro)$. 

The choice of $\MicroProbFwd$ is not compatible with our original derivation of the fluctuation theorem (\ref{eq:fluc2}), which assumes that it is everywhere nonzero. But we can recover a sensible expression if we restrict our domain of integration to trajectories that begin in $\MacroI$ and end in $\MacroII$ \cite{Vaikuntanathan2009,Berut2015}. This yields:
\begin{align}
\label{eq:flucirr}
\langle e^{-(\Delta \EntropyExt+\Delta\EntropyInt)/\Boltz } \rangle_{\MacroI\to\MacroII} = e^{\ln \frac{\pi(\MacroII\to\MacroI)}{\pi(\MacroI\to\MacroII)}}
\end{align}
where $\pi(\MacroI\to\MacroII)$ is the normalization of the trajectory distribution over the restricted domain
\begin{align}
\pi(\MacroI\to\MacroII) = \int_{\MacroI\to\MacroII} \Measure{\Micro_0^\EndTime} \MicroProbFwd\TrajProbControl
\end{align}
and gives the total probability of arriving in $\MacroII$ after time $\EndTime$ given that the system started in $\MacroI$.

The inequality $e^x \geq 1+x$ then gives \cite{England2013}:
\begin{align}
\label{eq:secondlawgen}
\langle \Delta\EntropyInt\rangle_{\MacroI\to\MacroII} + \langle \Delta \EntropyExt \rangle_{\MacroI\to\MacroII} \geq  - \Boltz \ln \frac{\pi(\MacroII\to\MacroI)}{\pi(\MacroI\to\MacroII)}.
\end{align}

This is a tighter version of the Clausius inequality (\ref{eq:StochClausius}) for the case where $\pi(\MacroI\to\MacroII)>\pi(\MacroII\to\MacroI)$. Generating internal order with a negative $\langle\Delta\EntropyInt\rangle$ requires dissipating \emph{at least} the same amount of entropy into the environment, but this minimum can only be achieved when the system is just as likely to return back to $\MacroI$ as it was to arrive in $\MacroII$ from $\MacroI$.

In Equation (\ref{eq:ShannonChange}) of the previous section, we noted that $\langle \Delta\EntropyInt\rangle$ is simply the change in the Shannon entropy of the distribution over states of the system. The restricted average appearing in Equation (\ref{eq:secondlawgen}), however, has a more subtle interpretation. Using Bayes' Rule, we can express this quantity in terms of the conditional distribution $p_0(\Micro_0|\to\MacroII)$ of finding the system in $\Micro_0$ at time 0 given that it ends up in $\MacroII$ at time $\EndTime$:
\begin{align}
\label{eq:ShannonChangeRest}
\langle \Delta \EntropyInt \rangle_{\MacroI\to\MacroII} =  -\Boltz &\int_\MacroII d\Micro \,\MicroProbRev \ln \MicroProbRev \nonumber\\
&+ \Boltz \int_\MacroI d\Micro \,p_0(\Micro_0|\to\MacroII) \ln \MicroProbFwd.
\end{align}
The first term on the right hand side is the ordinary Shannon entropy of the final state, but the second term is a cross-entropy between the actual initial distribution and the new distribution conditioned on the final macrostate. The thermodynamic interpretation of $\langle \Delta \EntropyInt \rangle_{\MacroI\to\MacroII} $ as a change in internal entropy requires the additional assumption that the macrostate dynamics on the timescale of interest are effectively Markovian, so that the probability of a transition from $\MacroI$ to $\MacroII$ in time $\EndTime$ is independent of the exact initial microstate within $\MacroI$.

For a macroscopic system, even down to the level of a single biological cell, the relative probability of the reverse transition $\pi(\MacroII\to\MacroI)/\pi(\MacroI\to\MacroII)$ quickly becomes astronomically small. If $\MacroI$ represents the state with a single bacterium on an agar plate, and $\MacroII$ is the state with two bacteria, then the probability of the transition $\MacroI$ to $\MacroII$ in twenty minutes can be close to 1, but the probability of the reverse transition $\MacroII \to \MacroI$ has been estimated at $\exp(-10^{11})$ \cite{England2013}. Even though this number is so small, Equation (\ref{eq:secondlawgen}) does not allow us to set it to zero: that would make the right-hand side infinite, and require an infinite increase in entropy to maintain the inequality. The above estimate already sets the right-hand side to $10^{11}\Boltz$, which corresponds to significant heat production at 300 Kelvin. For a liter of exponentially growing bacteria at a reasonable density of $10^6$/mL, half a Joule of heat per 30-minute division cycle must be dissipated even in the absence of a decrease in internal entropy. 

The original form of the Clausius relation (\ref{eq:StochClausius}) requires only that the heat dissipation compensate the change in internal entropy, which has been estimated at $-10^{10} \Boltz$ \cite{England2013}. This suggests that most of the heat generated by the bacterium during division goes into providing statistical irreversibility to the many intermediate steps of the process, ensuring that the key reactions reliably proceed in the correct direction.


\subsection{Application to Self-Assembly}
\label{sec:assembly}
We now turn to a set of systems that can be explicitly modeled, and provide a framework for obtaining design principles from Equation (\ref{eq:secondlawgen}). Consider a self-assembly process in which particles at chemical potential $\mu$ stick together to build up a structure. We parameterize the state of the assembly by the number of particles $N$ it contains and a set of intensive parameters $\omega$ that characterize the internal structure and composition. If $\mu$ is large enough, the structure will spontaneously grow at a finite rate. One can now examine the thermodynamic constraints on achieving a given probability distribution over $\omega$ that differs from the Boltzmann distribution. Of particular interest are probability distributions that increase the yield of a desired structure, as occurs in kinetic proofreading situations. This problem has been studied by several authors in the context of a simple but surprisingly rich model of template-based replication \cite{Andrieux2008,Andrieux2009,Esposito2010d,Sartori2013}.

 In this section, we use the tightened Clausius equality of Equation (\ref{eq:secondlawgen}) to re-derive two recently obtained general thermodynamic bounds on the dissipative cost of a non-Boltzmann distribution over $\omega$ \cite{Nguyen2016}. We first note that the environmental entropy change over a transition from $N$ to $N+\Delta N$ is given by (\ref{eq:entropyext}):
\begin{align}
\langle \Delta\EntropyExt \rangle_{N \to N+\Delta N} = \frac{\mu \Delta N - \langle\Energy\rangle_{N+\Delta N} +  \langle\Energy\rangle_N}{\Temp}
\end{align}
where $\langle \Energy \rangle_N$ is the average energy at fixed $N$ in the growing structure. 

To express the internal entropy change, we write the nonequilibrium distribution over microstates at fixed $N$ in terms of an effective energy and free energy:
\begin{align}
p_N(\omega) = e^{-\frac{1}{\Boltz\Temp}[\Energy^{\rm eff}(N,\omega) - \Free^{\rm eff}_N]}
\end{align}
where $\Free^{\rm eff}_N = -\Boltz\Temp\ln \sum_{\omega}e^{-\frac{\Energy^{\rm eff}(N,\omega)}{\Boltz\Temp}}$. Increasing the yield of a given structure $\omega$ is now equivalent to decreasing the relative $\Energy^{\rm eff}(N,\omega)$ for that $\omega$. A few lines of algebra confirm that 
\begin{align}
\langle \EntropyInt\rangle_N = -\Boltz\sum_{\omega}p_N(\omega)\ln p_N(\omega) = \frac{1}{\Temp} \left( \langle \Energy^{\rm eff}\rangle_N - \Free^{\rm eff}_N\right)
\end{align}
in this parameterization. 

In the large $N$ limit, we expect that the intensive quantity $\omega$ should be an average of identically distributed independent random variables, and so the quantities $\Free^{\rm eff}_N, \Free_N, \langle \Energy^{\rm eff}\rangle_N, \langle \Energy\rangle_N$ appearing in the exponent of the probability distributions should all become proportional to $N$ (cf. \cite{Touchette2009}). This implies that
\begin{align}
\frac{\Free^{\rm eff}_{N+\Delta N} - \Free^{\rm eff}_N}{\Delta N} = \frac{\Free^{\rm eff}_N}{N}
\end{align}
and likewise for the other quantities. 

In this case, Equation (\ref{eq:secondlawgen}) implies:
\begin{align}
\label{eq:Nguyen}
\delta \mu - &\frac{(\langle \Energy\rangle_N - \Free_N) - (\langle \Energy^{\rm eff}\rangle_N - \Free^{\rm eff}_{N}) }{N}\nonumber\\
&\,\,\,\,\,\,\,\,\,\,\,\,\,\,\,\,\,\,\,\,\,\,\geq \frac{\Boltz\Temp}{\Delta N}\ln \frac{\pi(N \to N+\Delta N)}{\pi(N+\Delta N \to N)} 
\end{align}
where $\delta\mu \equiv \mu - \Free_N/N$ is the excess chemical potential beyond what is required to keep $N$ stationary in the equilibrium ensemble. The left hand side has been rearranged so that the second term is the relative entropy between the actual distribution and the Boltzmann distribution (divided by $N$), which is always positive. Note that both averages are taken in the actual nonequilibrium distribution over microstates at fixed $N$. 

As in the self-replication example, making predictions using this result requires computing the reversal probability $\pi(N+\Delta N \to N)$. The first thing we can say about this quantity is that it is less than $\pi(N\to N + \Delta N)$ if $\Delta N$ is positive and the structure is growing, which makes the right hand side of (\ref{eq:Nguyen}) greater than zero. The resulting inequality is one of the main conclusions of the paper by Nguyen and Vaikuntanthan \cite{Nguyen2016}, and extends an earlier result by the latter author on systems driven by externally varied parameters \cite{Vaikuntanathan2009}. We have obtained it by a different route, which sheds light on the implications of the tightened Clausius inequality (\ref{eq:secondlawgen}). 

\begin{figure}
	\includegraphics[width=8.5cm]{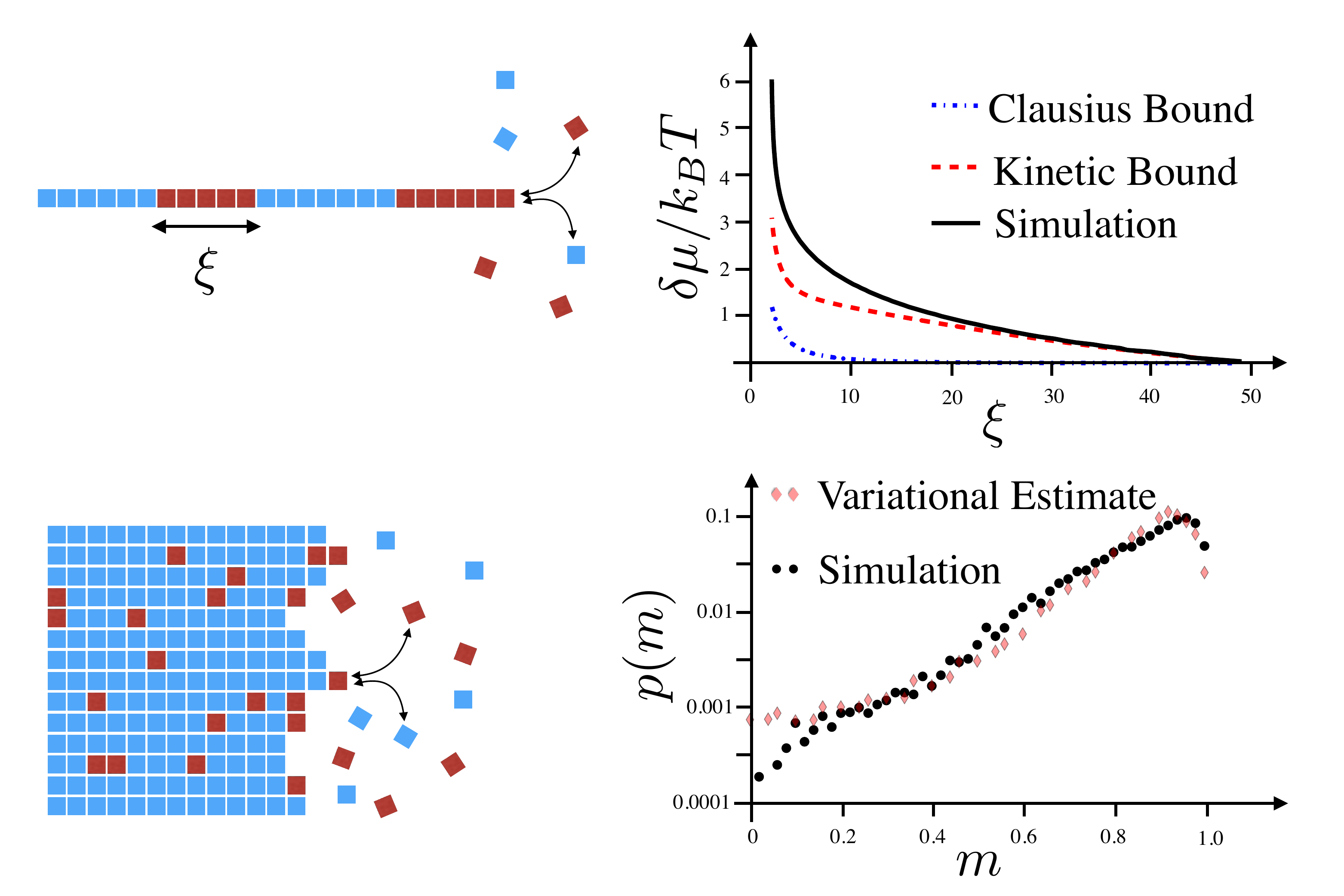}
	\caption{Color online. Plots generated using simulation data originally reported in \cite{Nguyen2016}, with permission from the authors. Red and blue monomers are maintained at identical chemical potentials $\mu$ in a supersaturated solution, but like-colored pairs of monomers bind more strongly than unlike pairs. Top: Excess chemical potential required to reduce the typical size $\xi$ of uniformly colored domains from its equilibrium value $\xi = 50$ in simulated assembly of a 1-dimensional filament. The black line is generated directly from simulation, while the blue line is obtained from Equation (\ref{eq:Nguyen}) by requiring that the left hand side be non-negative. The red line is the tighter bound of Equation (\ref{eq:Nguyen2}), which includes kinetic information about the relative size of fluctuations in assembly speed. Bottom: Probability of observing a fraction $m$ of blue monomers in a 2-dimensional filament that has been growing for a long time. The black dots are direct simulation data. The red diamonds are generated by sampling from an Ising distribution with coupling strength chosen to saturate the constraint (\ref{eq:Nguyen2}).}
	\label{fig:assembly}
\end{figure}

The top panel of Figure \ref{fig:assembly}, generated with data from \cite{Nguyen2016}, shows the minimum allowed $\delta \mu$ as a function of the characteristic size $\xi$ of spatial correlations in a 1-D filament assembled from two different monomer types. $E^{\rm eff}$ for a given $\xi$ is found by inverting the expression for $\xi$ as a function of binding energies at thermal equilibrium. Actual values of $\xi$ obtained by growing the filament at finite $\delta \mu$ in simulation satisfy this weakened version of (\ref{eq:Nguyen}) that simply requires the left-hand side to be positive. But there is plenty of room to obtain a stronger bound that makes use of the right hand side.

The statistical irreversibility can be calculated explicitly in terms of a measurable drift velocity $v$ and diffusion coefficient $D$ if the dynamics of $N$ are exactly described by a Gaussian drift-diffusion process. When inserted into the full expression for the irreversibility, this yields a simple ratio of $v$ to $D$ regardless of the exact value of $\Delta N$:
\begin{align}
\frac{1}{\Delta N}\ln \frac{\pi(N \to N+\Delta N)}{\pi(N+\Delta N \to N)} = \frac{v}{D}.
\end{align}

Outside of this special scenario, we can still define a drift velocity and diffusion coefficient in the limit $\EndTime, \Delta N \to \infty$, where the central limit theorem guarantees that small fluctuations in $\Delta N$ become Gaussian. The statistics of these fluctuations about the mean value $\langle \Delta N\rangle = v\EndTime$ are fully determined by a diffusion coefficient $D = \frac{1}{2} \lim_{\EndTime\to\infty} \frac{{\rm var}(\Delta N)}{\EndTime}$ where ${\rm var}(\Delta N) = \langle(\Delta N)^2\rangle - \langle \Delta N\rangle^2$ is the variance in $\Delta N$ over the observation time $\EndTime$. But the return probability becomes exponentially small in this limit, and involves an extremely rare large fluctuation that is not covered by the central limit theorem. As we will discuss at more length in Section \ref{sec:uncertainty}, recent results in the theory of rare fluctuations in stochastic processes imply that the Gaussian extrapolation always underestimates the true irreversibility:
\begin{align}
\label{eq:uncertaintyassembly}
\frac{1}{\Delta N}\ln \frac{\pi(N \to N+\Delta N)}{\pi(N+\Delta N \to N)} \geq \frac{v}{D}.
\end{align}
This allows us to use the observable ratio $v/D$ on the right hand side of the tightened inequality of Equation (\ref{eq:Nguyen}), leading to the second main result in \cite{Nguyen2016}:
\begin{align}
\label{eq:Nguyen2}
\delta \mu - \frac{(\langle \Energy\rangle_N - \Free_N) - (\langle \Energy^{\rm eff}\rangle_N - \Free^{\rm eff}_{N}) }{N} \geq \frac{v}{D}.
\end{align}

Returning to the top panel of Figure \ref{fig:assembly}, we see that this revised bound is significantly tighter. As in the self-replication example, the energetic cost of the process is mainly determined by the statistical irreversibility. As the forcing increases from zero, all the chemical work initially goes into breaking time-reversal symmetry and biasing the process towards positive growth. Only as $\delta\mu$ increases beyond $\sim2\Boltz\Temp$ does the internal entropy change begin to make a significant contribution. 

Because the new bound can be saturated at finite distance from equilibrium, it can also be used to predict properties of the assembled structure without solving the dynamics. The bottom panel of Figure \ref{fig:assembly}, also made with data from \cite{Nguyen2016}, displays a prediction of the distribution over the fraction $m = N_B/N$ of blue monomers in a 2-dimensional structure. This prediction was obtained by assuming that the nonequilibrium statistics of these fluctuations are still described by an equilibrium Ising model, but with a modified effective coupling energy. This effective coupling was chosen to make the distribution as different as possible from the actual equilibrium distribution while still satisfying the bound contained in Equation (\ref{eq:Nguyen2}). The agreement between the prediction and the observed fluctuation spectrum is encouraging, and more work should be done to determine how far this procedure can be generalized. In principle, it could be a powerful way to obtain design principles for more complicated self-assembly scenarios that are challenging to simulate directly.

\section{Dynamical precision and constraints on rare fluctuations}
\label{sec:uncertainty}
The microscopic reversibility relation (\ref{eq:MicroRev}) relates entropy changes to statistical irreversibility. Results derived from this relation thus tend to be most useful when some information is available about the probabilities of trajectories that reverse the system's typical behavior. In the previous section we discussed the ideal case of perfectly Gaussian fluctuations, where the reversal probability is completely determined by drift and diffusion coefficients. But in general, computing the irreversibility requires evaluating the probabilities of exponentially rare events, which do not fall under the central limit theorem, and may bear no relation to the small fluctuations actually observed in a finite-time experiment. 

For this reason, the power of nonequilibrium results like those of Section \ref{sec:shannon} has been greatly augmented by advances in the mathematical theory of rare events in stochastic processes, known as Large Deviation Theory (see \cite{Touchette2009} for an accessible introduction to this field). In this section, we present the recently discovered bound underlying the constraint on compositional fluctuations of Equation (\ref{eq:Nguyen2}) \cite{Gingrich2016,Pietzonka2016c}. This result also places constraints on dynamical design goals via a ``thermodynamic uncertainty relation'', which assigns an energetic cost to the suppression of fluctuations in the speed of a cyclic process \cite{Barato2015}. We will end the section by pointing out a few applications in this area.

\subsection{Bound on rare current fluctuations}
\begin{figure}
	\includegraphics[width=8.5cm]{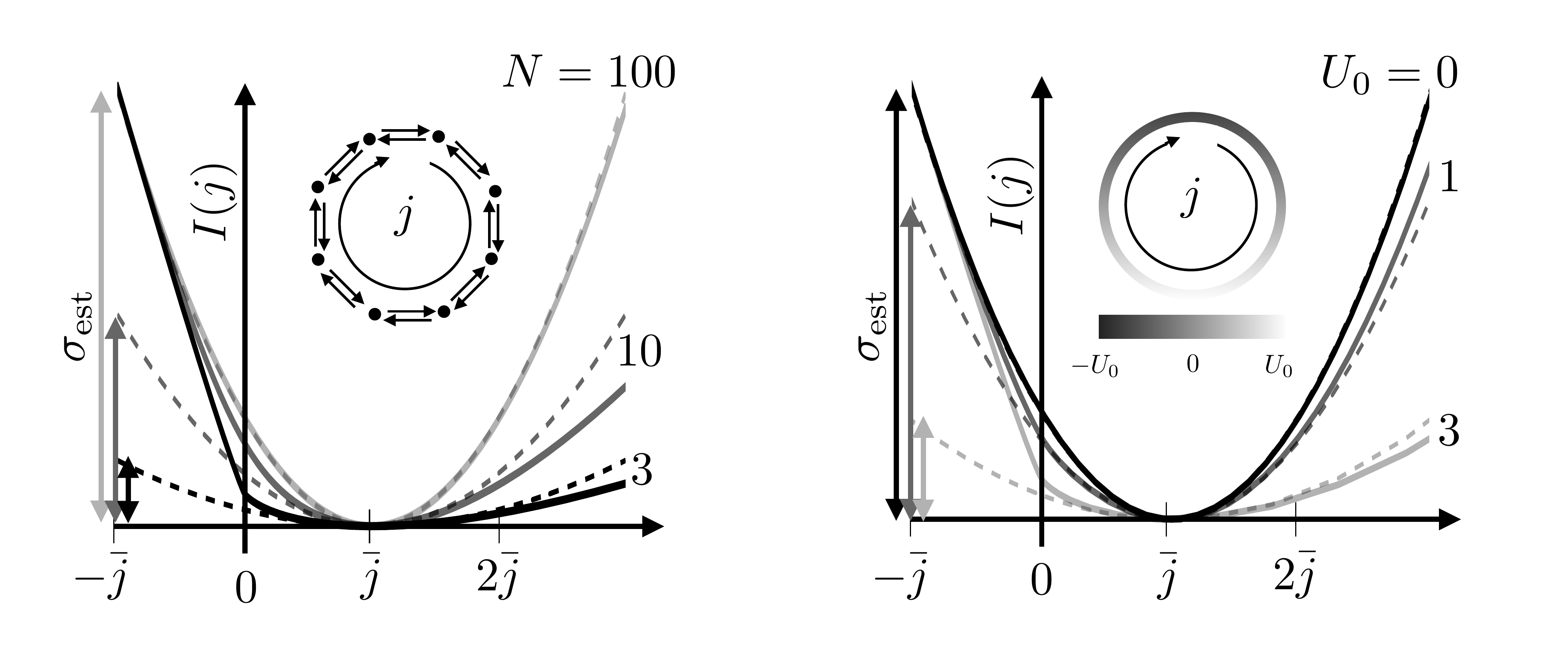}
	\caption{Large deviation rate function $I(j) = -\lim\frac{1}{\EndTime} \ln p_{\EndTime}(j)$ for a ring of $N$ discrete states, at fixed values of the typical current $\bar{j}$ and irreversibility $\sigma$. The solid lines are exact calculations, and dotted lines are Gaussian extrapolations $I_G(j)$ based on small fluctuations. Left: Uniform transition rates at three different $N$ values with $\sigma, \bar{j}$ fixed. The extrapolated probability of reversing the mean current is significantly different from the true $I(-\bar{j})$ for $N=3$, but improves as $N$ increases, until the lines overlap at $N = 100$. These overlapping lines also represent the quadratic upper bound $I_B(j)$ for these $\sigma,\bar{j}$ values. Right: Three different sets of non-uniform transition rates at fixed $N,\sigma,\bar{j}$. The number of states is fixed at $N = 100$, and the log-ratio of rates $\ln w_+/w_-$ varies sinusoidally with amplitude $U_0/N$. This could be used as a model of driven diffusion through the sinusoidal potential energy landscape shaded on the ring in the inset, with amplitude $U_0$ in units of $\Boltz\Temp$.  As $U_0$ becomes significantly larger than 1, the Gaussian estimate again begins to deviate downward from the true value of the entropy production rate.}
	\label{fig:ratefunction}
\end{figure}

 We illustrate the bound with a simple model: the ring of $N$ discrete states shown in the first panel of Figure \ref{fig:ratefunction}. The system can hop from one state to either adjacent state in a Poisson process, whose rates for clockwise and counterclockwise hops $w_+$ and $w_-$ are independent of the position along the ring. We can think of this as a highly simplified model of a biochemical clock, in which a protein cycles through a set of internal conformations or phosphorylation states \cite{Barato2016}. 
 
 Our goal is to constrain the statistical irreversibility of the typical dynamics based on small, observable fluctuations. Specifically, we can consider the probability $p_{\EndTime}(j)$ that the system completes $j\EndTime$ net cycles in the clockwise direction in time $\EndTime$. In the limit of large $\EndTime$, small fluctuations in $j$ are described by a Gaussian distribution peaked at the typical value (cf. \cite{vanKampen1981})
\begin{align}
\bar{j} &= \frac{1}{N}(w_+ - w_-)
\end{align}
with variance
\begin{align}
{\rm var}(j) &= \frac{1}{N\EndTime}(w_+ + w_-).
\end{align} 
We can quantify the statistical irreversibility with an expression analogous to the right hand side of the tightened Second Law bound in Equation (\ref{eq:Nguyen}), by taking the logarithm of the ratio of the probabilities of observing the system's typical behavior and its time-reverse: 
 \begin{align}
 \label{eq:statirr}
\sigma \equiv \lim_{\EndTime\to\infty} \frac{1}{\EndTime}\ln \frac{p_{\EndTime}(\bar{j} )}{p_{\EndTime}(-\bar{j})}.
\end{align}
Depending on the physical interpretation of the model, $\sigma$ may be equal to the rate of entropy production (in units of $\Boltz$). But for the moment, we are concerned only with the statistical properties of the model itself, abstracting from the thermodynamics, and have introduced this new symbol to reflect this distinction. In our simple ring model, the exact value of $\sigma$ can be found by examining the probabilities of the individual trajectories of $j\EndTime$ net positive steps, and noting that each one is related to the probability of its time-reversed version by the same factor $(w_+/w_-)^{Nj\EndTime}$. Inserting this into Equation (\ref{eq:statirr}) yields:
\begin{align}
\sigma = (w_+ - w_-) \ln \frac{w_+}{w_-}.
\end{align}
If we instead estimate the irreversibility based on small fluctuations, replacing $p_{\EndTime}(j)$ with a Gaussian distribution of mean $\bar{j}$ and variance ${\rm var}(j)$, we find:
\begin{align}
\label{eq:gauss1}
\sigma_{\rm est} = \frac{2}{\EndTime}\frac{\bar{j}^2}{{\rm var}(j)} = 2\frac{(w_+-w_-)^2}{w_+ + w_-}.
\end{align}
One can easily verify that $\sigma_{\rm est}\leq \sigma$ in this case, with equality only when $(w_+ - w_-)/(w_+ + w_-) \ll 1$. 

Barato and Seifert computed $\sigma$ and $\sigma_{\rm est}$ in a diverse array of more complex models, and found that the true irreversibility was always bounded by this simple function of ${\rm var}(j)$ and $\bar{j}$ \cite{Barato2015}. To prove this relation for a generic Markov process, Gingrich \emph{et al.} made use of recently discovered properties of the large deviation rate function $I(j)$, which represents the scaled logarithm of $p_{\EndTime}(j)$ in the long time limit:
\begin{align}
I(j) = -\lim_{\EndTime\to\infty} \frac{1}{\EndTime} \ln p_{\EndTime}(j).
\end{align}
This function is minimized at the most likely value of $j$, which is equal to the mean $\bar{j}$ in this limit. Since all the probability concentrates at this point, we always have $I(\bar{j}) = 0$. (See \cite{Touchette2009} for a mathematically precise construction of $I(j)$, accounting for the fact that $j$ on the right hand side must be an integer multiple of $1/\EndTime$, while $j$ on the left can be any real number). For large enough values of $\EndTime$, $I(j)$ plays a role analogous to free energy, and the distribution over $j$ is given by
\begin{align}
\label{eq:dist}
p_{\EndTime}(j) \propto e^{-\EndTime I(j)}.
\end{align}
The statistical irreversibility defined by Equation (\ref{eq:statirr}) can now be written as
\begin{align}
\label{eq:sigmaI}
\sigma = I(-\bar{j}).
\end{align}
And the estimate based on small fluctuations can be obtained from a quadratic expansion of $I(j)$ about $I(\bar{j})$:
\begin{align}
I_G(j) &\equiv \frac{1}{2}I''(\bar{j}) (j - \bar{j})^2.
\end{align}
Inserting this expression into Equation (\ref{eq:dist}) shows that the curvature of $I(j)$ controls the variance of the resulting Gaussian distribution via ${\rm var}(j) = 1/\EndTime I''(\bar{j})$. The estimated irreversibility from Equation (\ref{eq:gauss1}) can now be written in a parallel form to Equation (\ref{eq:sigmaI}):
\begin{align}
\label{eq:gauss2}
\sigma_{\rm est} = I_G(-\bar{j}).
\end{align}
We have plotted $I(j)$ and $I_G(j)$ in Figure \ref{fig:ratefunction} for various parameter values sharing the same $\sigma$ and $\bar{j}$ in two different models. The first is the uniform ring model presented above, and the second is a ring with non-uniform rates.

Gingrich \emph{et al.} showed that $I(j)$ is bounded from above by a different quadratic function which we will call $I_B(j)$ \cite{Gingrich2016}. $I_B(j)$ is uniquely defined by requiring that $I_B(\bar{j}) = I_B'(\bar{j}) = 0$ and $I_B(-\bar{j}) = \sigma$, with $\bar{j}$ and $\sigma$ taken from the original model. This bound was simultaneously discovered by Pietzonka, Barato and Seifert on the basis of numerical evidence \cite{Pietzonka2016c}. In an infinitesimal neighborhood around $j=\bar{j}$, we have $I_G(j) = I(j)$, and so $I_B(j)$ is also an upper bound on the approximate rate function $I_G(j)$ in this local region. But since $I_G(j)$ and $I_B(j)$ are quadratic functions tangent to each other at their extremal point $j=\bar{j}$, it is impossible for them to intersect, and the local bound implies a global bound on $I_G(j)$:
\begin{align}
\label{eq:Ibound}
I_G(j) \leq I_B(j).
\end{align}
Figure \ref{fig:ratefunction} illustrates this result for the current around a single ring of states. Both panels include one parameter set for which the true $I(j)$ is nearly quadratic, so that $I_B(j) \approx I_G(j) \approx I(j)$ ($N=100$ for left panel, $U_0 = 0$ for left panel). $I_B(j)$ is the same for all parameter sets within each panel, since it is fully determined by $\sigma$ and $\bar{j}$, and so this same curve can be used to compare $I_B(j)$ to $I(j)$ and $I_G(j)$ at different parameter values, revealing that $I_G(j)$ is always smaller than $I_B(j)$. Since $I_B(-\bar{j}) = \sigma$, this implies that $\sigma_{\rm est} = I_G(-\bar{j})$ always underestimates the true irreversibility:
\begin{align}
\label{eq:uncertainty}
\sigma_{\rm est}\leq \sigma.
\end{align}
These bounds continue to apply when the allowed transitions between states are not arranged on a single ring, but connected in an arbitrary topology containing any number of interlocking cycles \cite{Gingrich2016}. This makes the description of the dynamics more complicated, because the mean numbers of jumps per unit time for different transitions no longer need to be the equal. If $j_{xy}\EndTime$ is the net number of jumps from state $x$ to state $y$ observed in time $\EndTime$, then we can define a coarse-grained current $j_d$ by taking a linear combination of all these microscopic currents:
\begin{align}
j_d \equiv \sum_{y>x} d_{xy}j_{xy}
\end{align} 
where $d_{xy}$ are arbitrary constant coefficients. The statistical irreversibility $\sigma$ is now obtained as $\sigma = I(\{-\bar{j}_{xy}\})$ from the joint rate function $I(\{j_{xy}\})$ that depends on all the microscopic currents, and $I_B(j_d)$ is defined by $I_B(\bar{j}_d) = I_B'(\bar{j}_d) = 0$ and $I_B(-\bar{j}_d) = \sigma$. The Gaussian estimate is defined in the same way as in the single ring, using the mean and variance of $j_d$:
\begin{align}
\label{eq:gauss3}
\sigma_{\rm est} \equiv I_G(-\bar{j}_d ) =  \frac{2}{\EndTime}\frac{\bar{j}_d^2}{{\rm var}(j_d)}.
\end{align}
The rate function bound in Equation (\ref{eq:Ibound}) and the resulting bound on $\sigma$ in Equation (\ref{eq:uncertainty}) are purely mathematical results, derived from the properties of Markov processes prior to any thermodynamic interpretation. They provides new information that supplements the thermodynamic reasoning of Section \ref{sec:shannon}, converting formal expressions involving astronomically rare events into meaningful bounds on measurable quantities.

\subsection{Performance bounds in clocks, sensors and motors}
In Section \ref{sec:assembly} we saw how recent advances in nonequilibrium thermodynamics can tighten the Second Law bound on the cost of reducing the internal entropy of a thermodynamic system.  The bound on rare current fluctuations contained in Equation (\ref{eq:Ibound}) has spawned a new family of results with no parallel in thermal equilibrium, placing constraints on \emph{dynamical} performance goals. In these applications, the Gaussian estimate of irreversibility $\sigma_{\rm est}$ defined by Equation (\ref{eq:gauss3}) is itself the relevant figure of merit. Larger values of $\sigma_{\rm est}$ are associated with smaller relative fluctuations and more ``precise'' dynamics. If the Markov model under consideration adequately accounts for all the dissipative processes in the system, then microscopic reversibility (\ref{eq:MicroRev}) implies that the true statistical irreversibility $\sigma$ is equal to the steady-state rate of entropy change in the environment:
\begin{align}
\label{eq:sigmas}
\sigma = \frac{1}{\Boltz}\langle \dot{S}_e\rangle.
\end{align}
Equation (\ref{eq:uncertainty}) relating $\sigma$ and $\sigma_{\rm est}$ then specifies the minimum dissipative cost of achieving the desired precision, and is known as a ``Thermodynamic Uncertainty Relation,'' since its constraint on maximum precision is reminiscent of the Heisenberg Uncertainty Principle of quantum mechanics. In this subsection, we illustrate the consequences of this relation with some recent results on the precision of clocks and the efficiency of molecular motors.

Molecular motors like kinesin or the F1-ATPase convert chemical energy into mechanical work on small length scales where thermal fluctuations are important. The efficiency $\eta$ of a motor protein doing work against an applied force $\Force$ is defined as the ratio between the output work rate $\Force v$ and the mean input chemical work rate $\langle \dot{\WorkGen} \rangle$ in the steady state:
\begin{align}
\eta = \frac{\Force v}{\langle \dot{\WorkGen}\rangle}
\end{align} 
Here $v$ is the average speed of the motor traveling (or rotating) opposite to the direction of the force. The chemical work that is not transduced into mechanical work is released into the environment as heat, so we find:
\begin{align}
\langle \dot{\EntropyExt}\rangle &= \frac{1}{\Temp}\left(\langle \dot{\WorkGen}\rangle - \Force v\right)\\
&= \frac{\Force v}{\Temp} \left(\frac{1}{\eta} - 1\right).
\end{align}
By inserting this into the bound of Equation (\ref{eq:uncertainty}) and writing $\sigma_{\rm est}$ in terms of the mean speed $v$ and the diffusion coefficient $D$ of the motor motion, Pietzonka \emph{et al.} showed that the efficiency is constrained by \cite{Pietzonka2016a}:
\begin{align}
\eta \leq \frac{1}{1+v\Boltz\Temp/D\Force}.
\end{align}
Making the dynamics more deterministic by increasing the ratio $v/D$ thus incurs an extra energy cost, reducing the motor's efficiency.

Fluctuations are also a relevant design consideration in the construction of a clock, whose purpose is to provide a stable reference time for other processes. Any stochastic process that includes at least one cycle can be thought of as a clock, whose readout is the net number of forward jumps $j\EndTime$ across one of the links in the cycle. The bound on $\sigma_{\rm est}$ in Equation (\ref{eq:uncertainty}) then gives the minimum thermodynamic cost of a given precision goal, as discussed in \cite{Barato2016}, where the relative error $\epsilon$ is defined by
\begin{align}
\epsilon^2 \equiv \frac{{\rm var}(j)}{\bar{j}^2} = \frac{2}{\sigma_{\rm est}\EndTime}.
\end{align} 
The minimum entropy change in the environment $\langle \dot{S}_e \rangle\EndTime = \sigma\EndTime$ corresponding to a 1\% relative error, for instance, is $20,000\Boltz$. This bound is independent of $\EndTime$, so the \emph{rate} of entropy change $\langle\dot{S}_e\rangle$ depends on the timescale at which the precision is demanded. Since we are concerned with \emph{relative} error, higher precision is easier to obtain for longer time intervals. For instance, a 1\% error in the measurement of an hour using the stochastic clock only demands $20,000\Boltz$/hour, but the relative error in the measurement of each minute within that hour will be larger by a factor of $\sqrt{60}$. Achieving 1\% precision in the measurement of each minute would demand the much higher entropy production rate of $20,000\Boltz$/minute. As illustrated in Figure \ref{fig:ratefunction}, the maximum precision at fixed $\langle \dot{S}_e\rangle $ and $\bar{j}$ is obtained in the near-equilibrium regime of large $N$, where the entropy change per step $\Boltz \ln w_+/w_- = \sigma/N\bar{j}$ becomes small. 

This precision/dissipation trade-off had been investigated empirically before the bound of Equation (\ref{eq:uncertainty}) was discovered, in specific kinetic models of biological and artificial chemical clocks \cite{Cao2015}. The bound on $\sigma_{\rm est}$ puts these findings in a broader context, although it remains an open challenge to rigorously apply it to chemical concentration oscillations in an informative way. We can get an initial sense of the implications of the bound for these chemical clocks by replacing the full chemical reaction network with a phenomenological model like those of Figure \ref{fig:ratefunction}, where the system simply hops along a one-dimensional ring of $N$ states. In their original investigations, Cao \emph{et al.} expressed the clock precision in terms of fluctuations in the cycle period, whose average over many consecutive cycles can be related to the time-averaged current $j$ through a monitored link in our simple ring model as $\mathcal{T} = 1/j$ \cite{Cao2015}. An ensemble of initially synchronized clocks will drift apart due to thermal fluctuations, and each member will complete its cycle in a different period of time. The variance of the distribution over completion times will increase with each successive cycle, and the central limit theorem guarantees that this growth eventually becomes linear in the number of completed cycles $N_c = \bar{j}\EndTime$. This allows us to define a ``diffusion coefficient'' $D$ by
\begin{align}
{\rm var}(N_c \mathcal{T}) = D N_c \bar{\mathcal{T}} 
\end{align} 
where $\bar{\mathcal{T}} = 1/\bar{j}$ is the typical period. The dimensionless quantity $D/\bar{\mathcal{T}}$ measures the relative drift per cycle, and Cao \emph{et al.} found an empirical trade-off between this measure of the drift and the entropy production per cycle in the four models they studied \cite{Cao2015}.  

We can place a thermodynamic bound on $D/\bar{\mathcal{T}}$ by noting that the relationship between $\mathcal{T}$ and $j$ for small fluctuations in the $\EndTime\to\infty$ limit is adequately described by the linear approximation
\begin{align}
\mathcal{T} \approx \frac{1}{\bar{j}}\left(1 - \frac{j - \bar{j}}{\bar{j}}\right).
\end{align}
From this expansion, it follows that ${\rm var}(j)/\bar{j}^2 = {\rm var(\mathcal{T})}/\bar{\mathcal{T}}^2$, and so Equations (\ref{eq:gauss2}), (\ref{eq:sigmas}) and (\ref{eq:uncertainty}) imply that
\begin{align}
\label{eq:clockun}
\frac{D}{\bar{\mathcal{T}}} \geq \frac{2\Boltz}{\Delta S_c}
\end{align}
where $\Delta S_c \equiv \langle \dot{S}_e\rangle \EndTime/N_c$ is the steady-state entropy production per clock cycle. The findings of Cao \emph{et al.} are consistent with this result, but they cannot saturate the inequality at large $\Delta S_c$ because the number of states per cycle $N$ is a fixed feature of each model. As illustrated in Figure \ref{fig:ratefunction}, when the entropy change per reaction becomes larger than $\Boltz$, the Gaussian extrapolation begins to significantly underestimate the true irreversibility, which is here equal to the entropy production. At fixed $N$, it can be shown that $D/\bar{\mathcal{T}} \geq 1/N$ even in the limit $\Delta S_c\to \infty$ \cite{David1987}. 
This opens up an interesting question as to the factors that constrain the number of states per cycle. Larger values of $N$ lead to improved precision at fixed $\Delta S_c$, but other costs must eventually come into play, perhaps involving the absolute cycle speed or robustness against externally imposed perturbations. Including these additional factors in the analysis will allow for a tighter estimate of the maximum feasible efficiency.

\section{Thermodynamic potentials}
\label{sec:excess}
In the previous two sections, we have seen how the Second Law can be tightened to provide informative constraints on far-from-equilibrium processes. We now examine how far the traditional machinery of thermodynamics can be generalized to new kinds of states. In particular, we would like to look at those nonequilibrium states that most resemble equilibrium, with macroscopic properties that are independent of initial conditions and do not change in time. These include familiar situations like steady heat conduction, steady shear flow, and steady chemical flux through a chemostatted reaction network, as well as stationary states of novel materials like the contractile actin cytoskeleton of a living cell or a solution of self-propelled Brownian particles. 

In the absence of an externally maintained driving force, the steady state is characterized by a thermodynamic potential, such as the entropy $S$, the Gibbs free energy $G$ or the Helmholtz free energy $F$. This thermodynamic potential plays three distinct roles. First of all, the force required to modify a control parameter (such as the volume of a cylinder of gas) is given by a partial derivative of this function. Second, the values of quantities that are left free to fluctuate (like the pressure of a gas at fixed volume) are found by extremizing the function. Third, the thermodynamic potential obeys a Clausius relation, which defines the minimum amount of work or heat that must be exchanged with the system in order to accomplish a transition from one steady state to another. 

Much effort in nonequilibrium statistical mechanics over the past 75 years has been devoted to the search for generalizations of these potentials for nonequilibrium steady states (cf. \cite[pp. 51-54]{Lebon2008}). It is now clear that there is no function that simultaneously plays all three roles in a generic nonequilibrium system. But considerable progress has been made in formalizing a thermodynamic approach to each of these aspects individually. In the regimes where these approaches provide concrete predictions, most of their content was already anticipated in classical linear response theory. The new results can be seen as a way of systematizing and integrating those older findings, facilitating their application to more complex systems and possibly extending their range of validity.

A common thread running through all these efforts is the role of ``excess'' heat and ``excess'' work. Since a nonequilibrium steady state requires a continual supply of work and is constantly dissipating heat into its environment, the work and heat associated with a quasi-static transition or with relaxation to the steady state become infinite. To make use of these key thermodynamic quantities in the definition of a generalized thermodynamic potential, one must subtract off the ``housekeeping'' heat flow of the steady state, preserving only the excess portion associated with the given fluctuation or transition. 

In this section, we review the use of excess quantities in the calculation of statistical forces, the construction of variational principles, and the analysis of transitions between steady states. 

\subsection{Statistical Forces}
\label{sec:force}
Maes and coworkers have illustrated the role of excess work in near-equilibrium thermodynamics by investigating the thermodynamic forces $\Force_\Control$ conjugate to control parameters $\Control$ in isothermal steady states \cite{Basu2015,Basu2015a}. If we consider the isothermal compression of a nonequilibrium material, for instance, such as an active matter suspension or a sheared fluid, the control parameter $\Control$ will be the volume $V$ of the container, which we change by adjusting the position of one of the walls. The conjugate force $\Force$ is the pressure $P$ exerted by the material on that wall. In thermal equilibrium, the pressure is simply minus the derivative of the free energy: $P = -dF/dV$. The activity of the particles or the shear flow will modify this pressure, and we would like to know which aspects of this nonequilibrium activity determine the pressure change.

\begin{figure}
	\includegraphics[width=8.5cm]{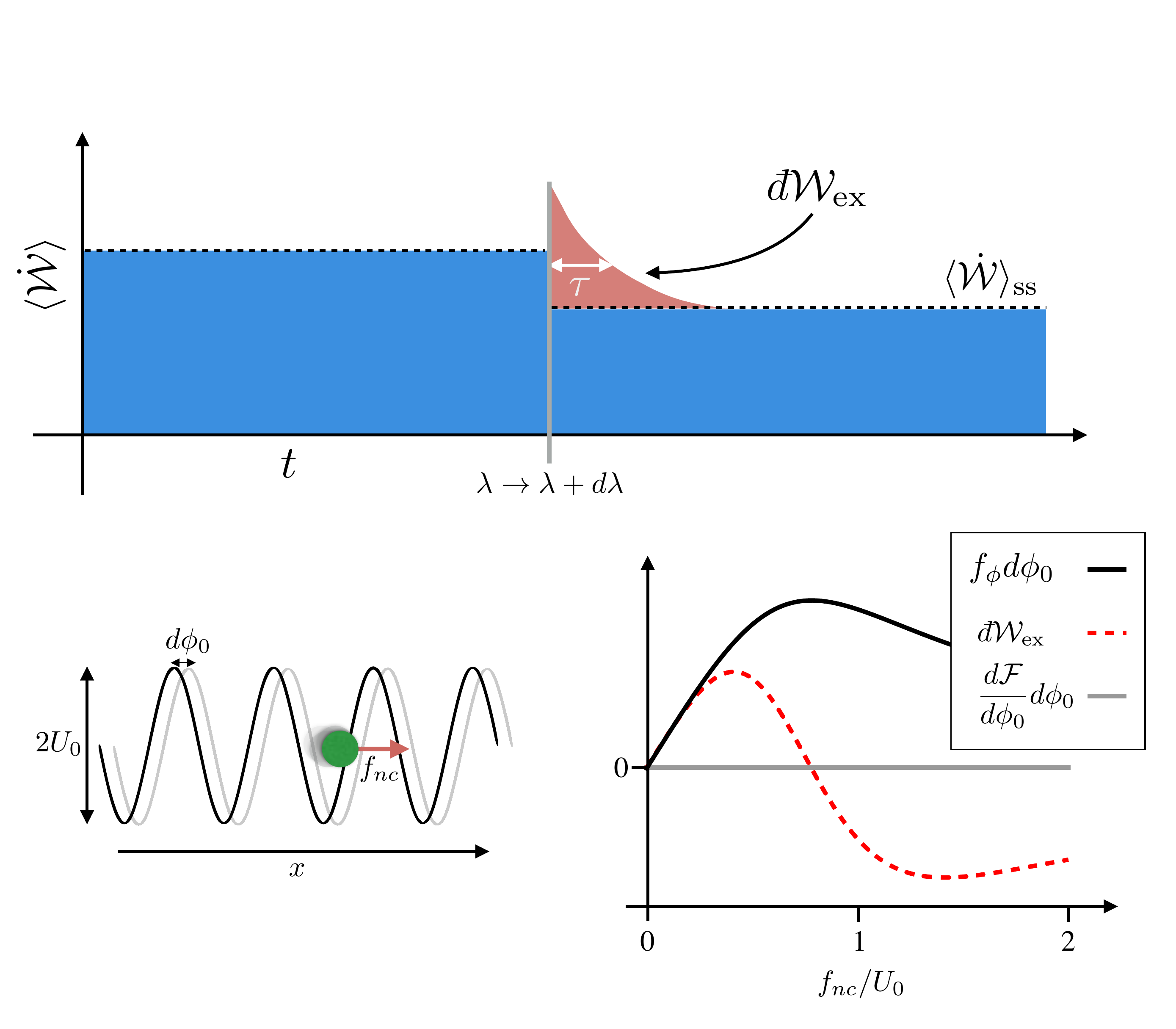}
	\caption{Color online. Top: Definition of excess work. When the control parameter $\Control$ is suddenly changed by a small amount $d\Control$, the average work rate $\langle \dot{\mathcal{W}}\rangle$ due to the steady driving forces is modified. The system relaxes to a new steady state on a time scale $\tau$. The excess work $\dbar \mathcal{W}_{\rm ex}$ associated with this transition is the difference between the actual work done during this relaxation and the work that would have been done if the system had started in the steady state. Left: Schematic of driven diffusion with periodic boundary conditions and a sinusoidal potential energy landscape of amplitude $U_0$. The phase $\phi_0$ of the periodic landscape can be used as a control parameter. Right: The work $-f_\phi d\phi_0$ required to quasi-statically change $\phi_0$ is equal to $-\dbar \mathcal{W}_{\rm ex}$ for small driving force $f_{nc}$. This guarantees that their difference is equal to $d\mathcal{F}=0$, as required by Equation (\ref{eq:ClausGen}). For larger driving, these two quantities diverge from each other, and Equation (\ref{eq:ClausGen}) is no longer satisfied.}
	\label{fig:diffusion}
\end{figure}

As a concrete example, consider the Brownian particle depicted in Figure \ref{fig:diffusion}, which diffuses in a one-dimensional ring with potential energy landscape $U(x) = U_0 \cos(x/2\pi -\phi_0)$ and periodic boundary conditions at $x=0$ and $x=1$. Our control parameter $\Control$ will be the phase $\phi_0$ of the landscape. Because of the symmetry of the system, the free energy $F$ is independent of $\phi_0$, and so the conjugate force $f^{\rm eq}_\phi = -dF/d\phi_0$ of equilibrium thermodynamics vanishes. Figure \ref{fig:diffusion} shows what happens when a non-conservative driving force $f_{nc}$ is turned on, which pushes the particle around the ring counterclockwise. The agent in control of $\phi_0$ now feels a nonzero average force $f_\phi$ trying to push $\phi_0$ in the direction of $f_{nc}$. 

Maes \emph{et al.} pointed out that the first-order nonequilibrium correction to the statistical force is determined by the excess work associated with a change in the corresponding parameter. As illustrated in Figure \ref{fig:diffusion}, the excess work $\dbar \WorkGen_{\rm ex}$ is the transient part of the work done by the steady nonequilibrium driving forces after a sudden change of parameters from $\Control$ to $\Control + d\Control$. To first order in the strength of the driving force, microscopic reversibility (\ref{eq:MicroRev}) requires that infinitesimal isothermal transitions satisfy an extended Clausius relation \cite{Basu2015,Komatsu2015}:
\begin{align}
\label{eq:ClausGen}
d\mathcal{F} = -f_\Control d\Control + \dbar \WorkGen_{\rm ex}
\end{align}
where $d\mathcal{F}$ is the change in the nonequilibrium free energy defined by Equation (\ref{eq:freegen}): $\FreeGen = \Energy - \Temp\langle \EntropyInt\rangle$. To linear order in the nonequilibrium drive strength, $d\mathcal{F}$ remains equal to the change in equilibrium free energy $dF$  \cite{Basu2015}. The right hand side of the equation is simply the total remaining work after subtracting the constant background of the steady-state work rate. Equation (\ref{eq:ClausGen}) thus says that quasi-static transitions between near-equilibrium steady states follow the Clausius relation $\Delta F = \WorkGen$ after this background subtraction. 

As illustrated in Figure \ref{fig:diffusion}, however, the fact that $\dbar \WorkGen_{\rm ex}$ is not an exact differential makes some processes possible that would be forbidden in classical thermodynamics. When $f_{nc} > 0$, the force $f_\phi$ conjugate to $\phi_0$ is positive all the way around the ring, and cannot be written as the gradient of any potential function. More generally, if multiple control parameters are allowed to vary, the integral of $f_\Control \cdot d \Control$ around a closed loop in control parameter space can be nonzero. This means that a cyclic process is possible that extracts work from the system, which would be a violation of the Second Law in classical thermodynamics. A nonequilibrium steady state relies on a constant supply of work even when the control parameters are not changing, and the right kind of change of parameters can extract some of this work. 

A more surprising consequence of Equation (\ref{eq:ClausGen}) for thermodynamic reasoning is that the statistical force depends on the \emph{kinetics} of the material. This stands in stark contrast to thermal equilibrium, where knowledge of microstate energies is sufficient to compute forces as derivatives of free energy. The source of this kinetic dependence is illustrated in the top panel of Figure \ref{fig:diffusion}, which shows how the excess work $\dbar \WorkGen_{\rm ex}$ done by the constant background driving forces depends on the time $\tau$ required to return to the steady state. This kinetic contribution enables boundary terms to significantly affect the pressure of macroscopic active materials, since the kinetics of relaxation after displacement of a boundary wall depend on the details of the interaction with that wall. Solon \emph{et al.} have recently demonstrated this for a model of active Brownian particles, where the mean force exerted by the suspension against a given wall generically depends on the stiffness of that wall, in a way that is independent of system size \cite{Solon2015b}. This implies that the pressure generally fails to be fully determined by other intensive parameters like concentration and temperature, but remains sensitive to the properties of the container even in the thermodynamic limit.

As the strength of the driving force increases, the first-order approximation in Equation (\ref{eq:ClausGen}) eventually breaks down. Figure \ref{fig:diffusion} shows how the excess work differential $\dbar \WorkGen_{\rm ex}$ remains well-defined for all values of the driving force, but need not bear any obvious relationship to $f_\Control$ or $d\mathcal{F}/d\Control$. It may still be a useful quantity for characterizing transitions between nonequilibrium steady states, however. Its behavior as a function of $f_{nc}$, for example, is very rich near the critical force $f_{nc} = U_0$, and may be related to other phenomena like giant diffusion and increased effective temperature that have been observed in this regime \cite{Reimann2001,Hayashi2004}.

\subsection{Variational Principles}
Thermodynamic potentials also provide variational principles for predicting the properties of systems at thermal equilibrium. Returning to the example of isothermal compression, we recall that the equilibrium probability $p_{\rm eq}(P)$ of a given fluctuation away from the deterministic pressure $P^*$ is determined by the free energy (cf. \cite{Einstein1910}):
\begin{align}
\label{eq:Einstein}
p_{\rm eq}(P) \propto e^{-F(P)/\Boltz\Temp}.
\end{align}
This relationship is essentially a coarse-grained version of the Boltzmann distribution. As the system size increases, the relative size of fluctuations decreases, and all the probability concentrates at the value $P = P^*$ that minimizes the free energy $F$.

\begin{figure}
	\centering
	\includegraphics[width=8.5cm]{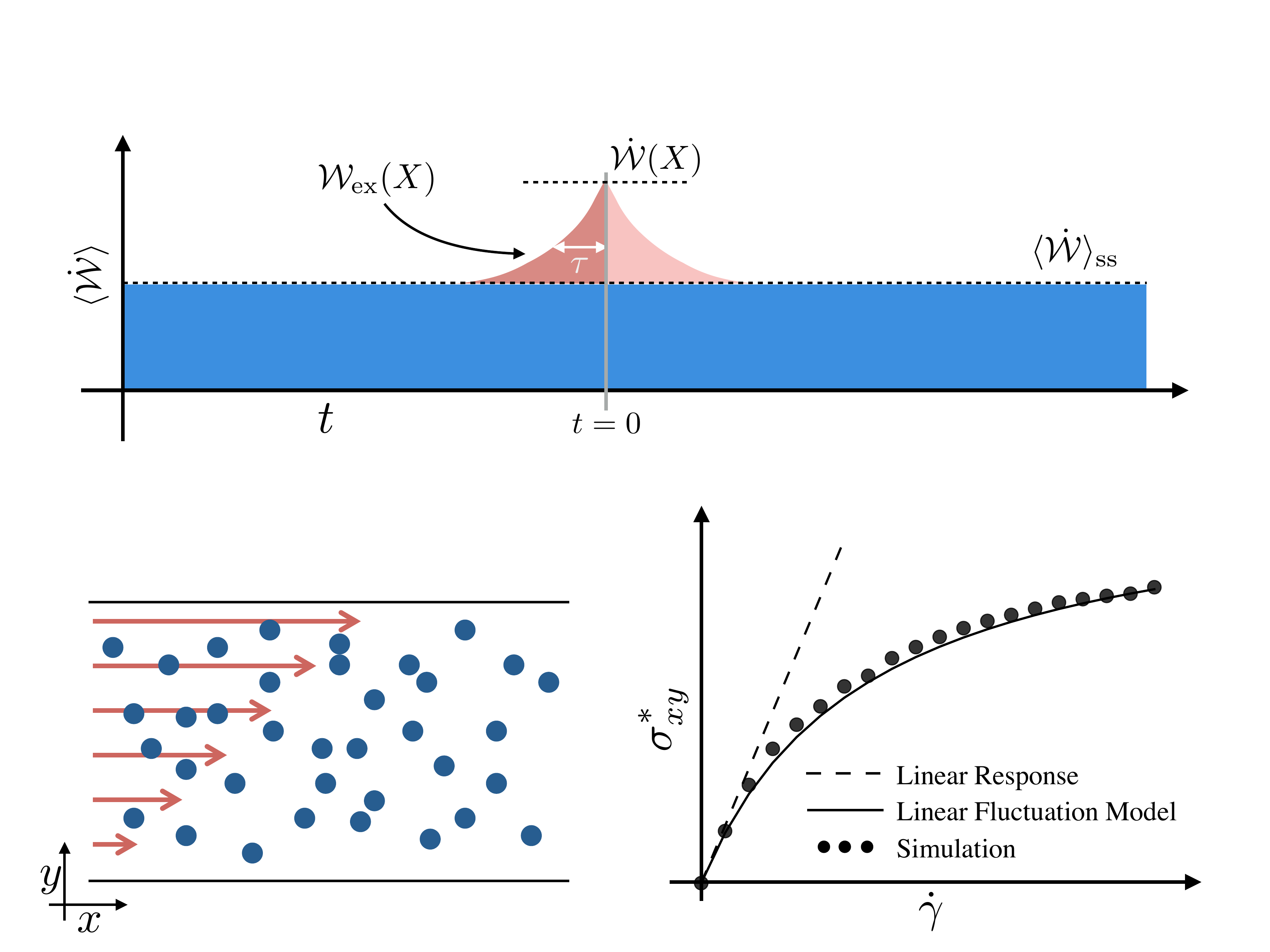}
	\caption{Color online. Top: Excess work associated with steady-state fluctuations. The mean steady-state work rate $\langle \dot{\WorkGen}\rangle_{\rm ss}$ is constant in time, but conditioning on a given fluctuation in an observable $X$ at time $t=0$ makes the average deviate from $\langle \dot{\WorkGen}\rangle_{\rm ss}$. The excess work $\WorkGen_{\rm ex}$ is the part of this deviation leading up to $t=0$.  Left: A nonequilibrium steady state can be set up in a colloidal suspension by imposing a fixed flow gradient $v_x = \dot{\gamma}y$ in the background solvent. Blue circles are Brownian particles interacting via a repulsive potential, and red arrows represent the imposed flow field. Right: The steady-state shear stress can be accurately predicted by minimizing $F - \WorkGen_{\rm ex}$ using the simple phenomenological model of stress fluctuations contained in Equations (\ref{eq:lineardyn}) and (\ref{eq:taumodel}).}
	\label{fig:Shear}
\end{figure}

In 1959, James McLennan showed that the correction to this distribution for near-equilibrium steady states is related in a simple way to the work done by the external driving forces \cite{McLennan1959,Maes2010}. The coarse-grained version of his finding gives:
\begin{align}
\label{eq:McLennan}
p_{\rm ss}(P) \propto e^{-[F(P) - \Work_{\rm ex}(P)]/\Boltz\Temp}
\end{align}
where $\Work_{\rm ex}(P)$ is the extra work done on the way to a given fluctuation in $P$, over and above the steady-state work rate, as illustrated in Figure \ref{fig:Shear}. In the linear-response limit considered by McLennan, the extra work on the way to a fluctuation is the same as the extra work during the relaxation of the fluctuation, and so the $\Work_{\rm ex}$ appearing here is essentially the same quantity discussed in the context of statistical forces above. Wherever Equation (\ref{eq:McLennan}) is valid, steady-state values of observables can be computed by minimizing $F - \Work_{\rm ex}$. By applying this variational principle, one can obtain the Green-Kubo linear response formula \cite{Maes2010,Komatsu2008,Komatsu2009} and other standard results of linear response theory.

The microscopic reversibility relation (\ref{eq:MicroRev}) provides a foundation for deriving exact generalizations of Equation (\ref{eq:McLennan}) for steady states arbitrarily far from equilibrium \cite{Crooks1999,Komatsu2008,Komatsu2009,Maes2010,Colangeli2011,Bertini2015}. These expressions involve quantities that are extremely challenging in general to measure or even estimate, and do not directly supply any new predictive power. But they provide a framework for systematically constructing perturbative expansions around the regime described by Equation (\ref{eq:McLennan}), shedding new light on the physical factors that set the limits of linear response theory. 

Following this line of reasoning, we have recently shown that the ``near-equilibrium'' variational principle contained in Equation (\ref{eq:McLennan}) can accurately describe driven steady states far outside the realm of classical linear response theory \cite{Marsland2015,MarslandThesis}. Our results complement those of Bunin \emph{et al.}, who showed that linear-response type results can remain valid under strong driving as long as the third cumulant of the work distribution remains sufficiently small \cite{Bunin2011,Bunin2013a}. Consider for example the sheared colloidal suspension illustrated in Figure \ref{fig:Shear}. The microstate of the system is defined by the positions of a set of interacting Brownian particles diffusing in a viscous solvent of volume $V$. The spatial configuration of the particles equilibrates on a timescale $\tau_0$ set by the diffusion coefficient, the number of particles, and the interparticle potential. A nonequilibrium steady state is created by moving the top of the container at a constant speed to set up a steady shear flow, where the solvent flow velocity in the $x$ direction is proportional to the $y$ position: $v_x = \dot{\gamma}y$. The constant of proportionality $\dot{\gamma}$ is called the shear rate, and is a measure of the nonequilibrium driving strength in this system. The instantaneous drag force per unit area $\sigma_{xy}$ resisting this movement is a function of the particle configuration, since the mutual repulsion of the particles can either assist or oppose the flow gradient depending on their relative positions. 

We showed that the variational principle of Equation (\ref{eq:McLennan}) correctly predicts the typical value of $\sigma_{xy}$ whenever its fluctuations are well described by a linear overdamped Langevin equation:
\begin{align}
\label{eq:lineardyn}
\frac{d}{dt}\sigma_{xy} = -\frac{1}{\tau} (\sigma_{xy} - \langle \sigma_{xy}\rangle_{\rm ss}) + \sqrt{2D}\xi(t)
\end{align}
where the Gaussian noise term is defined by $\langle \xi\rangle = 0$, $\langle \xi(t)\xi(t')\rangle = \delta(t-t')$, and $\tau,D$ are independent of $\sigma_{xy}$. The excess work in this model is easily computed in terms of $\sigma_{xy}$, $\dot{\gamma}$ and the relaxation time $\tau$:
\begin{align}
\WorkGen_{\rm ex}(\sigma_{xy}) = V\tau\dot{\gamma}(\sigma_{xy} - \langle \sigma_{xy}\rangle_{\rm ss}).
\end{align}
Minimizing $F - \WorkGen_{\rm ex}$ generates a prediction for the typical drag force $\sigma_{xy}^*$:
\begin{align}
\label{eq:sigmatyp}
\sigma_{xy}^* = \sigma_{xy}^0 + \frac{1}{\Boltz\Temp} \dot{\gamma}\tau V \langle \sigma_{xy}^2\rangle_{\rm eq},
\end{align}
where $\sigma_{xy}^0$ is the drag force in the absence of inter-particle repulsion, and is independent of the particle configuration. 

Near equilibrium, when the relaxation time $\tau$ is still equal to its equilibrium value $\tau_0$, Equation (\ref{eq:sigmatyp}) reproduces the Green-Kubo prediction of classical linear response theory. But at larger values of $\dot{\gamma}$, we expect the flow gradient to help the particle configuration relax more quickly. Eventually this convective relaxation becomes the dominant mechanism of particle rearrangement, and the relaxation timescale is roughly the time $1/\dot{\gamma}$ required for the flow gradient to push two adjacent particles past each other. We can capture both the $\dot{\gamma}\to 0$ and $\dot{\gamma} \to \infty$ limits by writing
\begin{align}
\label{eq:taumodel}
\tau = \frac{\tau_0}{1 + k\dot{\gamma}\tau_0}
\end{align}
where $k$ is a constant fitting parameter. Figure \ref{fig:Shear} compares the $\sigma_{xy}^*$ values directly obtained from a numerical simulation with the variational prediction of Equations (\ref{eq:sigmatyp}) and (\ref{eq:taumodel}). The simulation method and results are reported in more detail in \cite{Marsland2015,MarslandThesis}. The equilibrium relaxation time $\tau_0$ was obtained from the autocorrelation function of $\sigma_{xy}$ at thermal equilibrium, and $k=2$ was chosen to give the best fit to the rest of the curve. Also plotted is the linear response prediction using $\tau = \tau_0$, which only captures the initial slope of the $\sigma_{xy}$ vs. $\dot{\gamma}$ curve. 

We have generalized this procedure to encompass a broad class of systems and possible observables, as described in detail in \cite{MarslandThesis}. The excess work $\WorkGen_{\rm ex}$ provides the appropriate nonequilibrium correction to the equilibrium variational principle, as long as nonlinear corrections to the linearized fluctuation dynamics change $\WorkGen_{\rm ex}$ by much less than $\Boltz\Temp$ in typical fluctuations. But we also showed that the regime where Equation (\ref{eq:McLennan}) applies is precisely the regime where the nonequilibrium driving force merely adds a linear tilt to the effective free energy. $\WorkGen_{\rm ex}(X)$ is a linear function of the observable $X$, and if $F(X)$ is quadratic, then subtracting $\WorkGen_{\rm ex}$ shifts the location of the global minimum without changing the shape of the function. This suggests that thermodynamically meaningful variational principles may not apply to pattern-formation scenarios or other cases where new phase transitions appear that did not exist in the equilibrium. In this sense, the extended Green-Kubo relation of Equation (\ref{eq:sigmatyp}) is still a ``near-equilibrium'' result, since it only applies when the driving force incrementally modifies the equilibrium state, and fails when qualitatively new phenomena are generated.

\subsection{Clausius Inequality}
In Section \ref{sec:force}, we saw that quasi-static processes near equilibrium satisfy the generalized form of the Clausius relation expressed in Equation (\ref{eq:ClausGen}). For transitions between equilibrium states, an even stronger statement can be made using the Clausius inequality obtained in Equation (\ref{eq:StochClausiusWork}):
\begin{align}
\label{eq:ClausEq}
\WorkGen \geq \Delta \FreeGen,
\end{align}
where $\FreeGen = \Energy - T\langle \EntropyInt\rangle$ becomes equal to the equilibrium free energy $F$ in the absence of driving forces. Our derivation of Equation (\ref{eq:ClausEq}) from the fluctuation theorem in Section \ref{sec:shannon} remains valid arbitrarily far from equilibrium, but the relation is not very informative for transitions between nonequilibrium steady states. Work is constantly being done even in the steady state to keep the system out of equilibrium, so $\WorkGen\to\infty$ in the quasi-static limit and the inequality is trivially satisfied. 

A more interesting quantity is the finite ``applied'' work 
\begin{align}
\WorkGen_{\rm app} = \int_0^\EndTime \left \langle \frac{\partial \Energy}{\partial \Control}\right \rangle_t  \dot{\Control} \,dt
\end{align}
 done by the external agent who manipulates the control parameters $\Control$. The average is over the probability distribution at time $t$, obtained by evolving the initial distribution under the corresponding Fokker-Planck or Master equation. When the parameters are changed sufficiently slowly, the distribution is always very close to the steady state, and $\WorkGen_{\rm app}$ is simply minus the integral of the steady-state force $\Force_\Control$ discussed in Section \ref{sec:force}. A na\"{i}ve observer ignorant of the internal forces or gradients sustaining the nonequilibrium steady states would believe that $\WorkGen_{\rm app}$ is simply equal to the total work, and would expect it to satisfy Equation (\ref{eq:ClausEq}).
 
As we saw in Section \ref{sec:force}, this expectation can already fail in the linear response regime. In the example of Figure \ref{fig:diffusion}, $d \FreeGen = 0$, but $\dbar \WorkGen_{\rm app} = -f_\phi d\phi_0$ is negative. To gain a deeper understanding of why the equilibrium predictions fail and how they should be corrected, it would be useful to have a modified version of the Clausius inequality for $\WorkGen_{\rm app}$, in which equality can be achieved for quasi-static transitions. Hatano and Sasa showed in 2001 that this relation exists for systems described by overdamped Langevin dynamics \cite{Hatano2001}, and their findings were later extended to Poisson processes \cite{Esposito2010} and fluctuating hydrodynamics \cite{Bertini2015}. It is now clear that a whole family of inequalities can be obtained using arguments very similar to our derivation of the original Clausius inequality in Section \ref{sec:shannon}. 

The key insight behind these results lies in the fact that the trajectory probabilities $\RevState{\mathcal{P}}[\Rev{\Micro}_0^\EndTime]$ of the reverse process are integrated out when moving from the microscopic reversibility relation (\ref{eq:MicroRev}) to the integral fluctuation theorem (\ref{eq:fluc1}) \cite{Esposito2010,Seifert2012}. This means that integral fluctuation theorems for other quantities can be obtained by replacing the reverse trajectory probability with another properly normalized distribution over trajectories $\mathcal{P}^\dagger[\Rev{\Micro}_0^\EndTime]$. In this way, an integral fluctuation theorem $\langle e^{-\mathcal{A}} \rangle = 1$ can be obtained for any trajectory functional $\mathcal{A}[\Micro_0^\EndTime]$ that can be written in the form
\begin{align}
\mathcal{A}[\Micro_0^\EndTime] = - \ln \frac{\mathcal{P}^\dagger[\Rev{\Micro}_0^\EndTime]}{\mathcal{P}[\Micro_0^\EndTime]}
\end{align}
for some choice of $\mathcal{P}^\dagger[\Rev{\Micro}_0^\EndTime]$. As noted in Section \ref{sec:shannon}, the inequality $e^x \geq x+1$ then implies that $\langle \mathcal{A}\rangle \geq 0$, 

Hatano and Sasa identified a quantity that can be written in this form, reduces to the entropy change when the driving forces are turned off, and vanishes in a nonequilibrium steady state. This quantity is now known as the ``non-adiabatic'' entropy change $\Delta S_{\rm na}$ \cite{Esposito2010}. It is closely related to two other quantities known in the literature as the ``excess heat'' \cite{Hatano2001,Esposito2010} and ``excess work'' \cite{Bertini2015}. These reduce to the heat and work, respectively, for systems whose steady state obeys detailed balance, and they exactly satisfy modified Clausius relations. But they do not correspond in general to any measurable physical property \cite{Bertini2015,Sasa2014}. For this reason, we reserve the symbol $\WorkGen_{\rm ex}$ in this review for the physically meaningful excess work found by subtracting the (measurable) average steady state work rate from the actual work done by the continually acting nonequilibrium driving forces. But the fluctuation theorem for $\Delta S_{\rm na}$ still helps us understand how nonequilibrium steady states differ from equilibrium states. It turns out that $\Delta S_{\rm na}$ can be written in terms of derivatives of the trajectory-level entropy $\EntropyInt = - \Boltz \ln p_{\rm ss}$. For systems described by overdamped Langevin dynamics (see Section \ref{sec:models}), this allows us to rewrite $\langle \Delta S_{\rm na} \rangle \geq 0$ as 
\begin{align}
\WorkGen_{\rm app} \geq \Delta \FreeGen - \int_0^\EndTime \left\langle \nabla (\Energy-\Temp\EntropyInt) \cdot \dot{\Micro} \right\rangle_t \, dt.
\end{align}
This reduces to the ordinary Clausius inequality for undriven systems where $p_{\rm ss} \propto \exp(-\Energy/\Boltz\Temp)$, since then $\Temp\nabla\EntropyInt = \nabla \Energy$ and $\FreeGen = \Free$. But when the nonequilibrium steady-state distribution becomes different from the equilibrium distribution, new possibilities open up. It remains the case that equality is achieved only in the limit of quasi-static variation of $\Control$. But unlike the work-free energy theorem of classical thermodynamics, this result contains path-dependent quantities on both sides of the inequality. Thus the quasi-static work no longer defines a state function, and there is no longer any guarantee that the slow protocol extracts the maximum amount of work for a given path in control-parameter space. 

\section{Discussion}
\label{sec:distance}
The Second Law of Thermodynamics requires that the total entropy of the system plus environment can never decrease: so if the entropy of the system decreases, this must be compensated by the flow of heat into the environment. But in Section \ref{sec:shannon} we saw that the microscopic reversibility relation also attaches an entropic cost to statistical irreversibility, leading to much higher levels of entropy production than would have been demanded by the Second Law alone. When the entropic requirements of statistical irreversibility are accounted for, the remaining dissipation can provide a remarkably tight bound on the allowed internal entropy. In the example of compositional fluctuations in two-dimensional self-assembly, the bound was tight enough to provide an accurate estimate of the distribution of fluctuations without any other knowledge of the system dynamics. 

For many systems of interest -- especially those with many degrees of freedom -- the statistical irreversibility is difficult to compute or even estimate, since it involves the probability of extremely rare reverse trajectories. The self-assembly calculations of Section \ref{sec:shannon} relied on a recent result from Large Deviation Theory that relates the irreversibility to a measure of the size of typical fluctuations, which can be readily ascertained in an experiment or simulation. In Section \ref{sec:uncertainty}, we described how this result guarantees that a na\"{i}ve extrapolation of Gaussian fluctuations in a time-averaged probability current always underestimates the true irreversibility. We also discussed the new constraints it imposes on dynamical design goals, where the size of typical current fluctuations is itself the relevant figure of merit. 

In a nonequilibrium steady state, thermodynamic potentials no longer provide shortcuts for calculating statistical forces, predicting typical values of fluctuating observables, or constraining the minimum amount of heat or work required to bring the system from one state to another. Efforts to identify nonequilibrium thermodynamic potentials have shed considerable new light on the reasons why these shortcuts fail, and on the special features of thermal equilibrium that make them work. We saw in Section \ref{sec:excess} that empirically meaningful generalizations can be obtained close to equilibrium, where the thermodynamic quantities are modified by adding or subtracting the ``excess'' heat or work associated with a transition or fluctuation. This provides a powerful unified perspective in which to justify and organize the results of classical linear response theory, facilitating their application to new situations and extending their range of validity. 

In summary, it appears that nonequilibrium thermodynamics is finally approaching full maturity. While some of the early enthusiasm for universal principles has been disappointed, rigorous investigation has revealed an even richer structure whose outlines are now becoming clear. This theoretical framework is now in a position to provide guidance for specific applications in biotechnology and materials science, and feedback from real design goals in these areas will be crucial for deepening and solidifying our understanding. 

\section*{Acknowledgments}
RM acknowledges Government support under and awarded by DoD, Air Force Office of Scientific Research, National Defense Science and Engineering Graduate (NDSEG) Fellowship, 32 CFR 168a. J. E. thanks the Cabot family for their generous support of MIT faculty, and the James S. McDonnell Foundation for Scholar Grant No. 220020476. The authors thank J. Horowitz, G. Bisker and T. Gingrich for helpful comments.

\bibliographystyle{abbrv}
\bibliography{library}
\end{document}